\documentclass[aps, prb, amsmath, amssymb, twocolumn]{revtex4}
\usepackage{algorithm, algorithmic, graphicx, graphics, subfigure, times, caption, setspace} 
\usepackage{color}
\begin{document}

\title{A practical approach to Hohenberg-Kohn maps based on many-body
  correlations: learning the electronic density}
%\title{A practical approach to bypassing Kohn-Sham density functional
%  theory for three-dimensional systems using Hohenberg-Kohn maps}

\author{Edgar Josu\'e Landinez Borda}
% \email{landinez@}
\affiliation{Physics Division, Lawrence Livermore National Laboratory,
  Livermore, California 94550, USA} 

\author{Amit Samanta}
\email{samanta1@llnl.gov}
\affiliation{Physics Division, Lawrence Livermore National Laboratory,
  Livermore, California 94550, USA} 

\date{\today}

\begin{abstract}
High throughput screening of materials for technologically relevant areas,
like identification of better catalysts, electronic materials, ceramics
for high temperature applications and drug discovery, is an emerging topic
of research. To facilitate this, density functional theory based (DFT)
calculations are routinely used to calculate the electronic structure of
a wide variety of materials. However, DFT calculations are expensive and
the computing cost scales as the cube of the number of electrons present
in the system. Thus, it is desirable to generate surrogate models that
can mitigate these issues. To this end, we present a two step procedure
to predict total energies of large three-dimensional systems (with periodic
boundary conditions) with chemical accuracy (1 kcal/mol) per atom using a small data
set, meaning that such models can be trained on-the-fly. Our procedure is
based on the idea of the Hohenberg-Kohn map proposed by Brockherde et al.
(Nat. Commun, {\bf 8}, 872 (2017)) and involves two training models: one,
to predict the ground state charge density, $\rho \left( {\bf r} \right)$,
directly from the atomic structure, and another to predict the total energy
from $\rho \left( {\bf r} \right)$. To predict $\rho \left( {\bf r} \right)$,
we use many-body correlation descriptors to accurately describe the
neighborhood of a grid point and to predict the total energy we use
amplitudes of these many-body correlation descriptors. Utilizing the amplitudes
of the many-body descriptors
% to predict the total energy
allows for uniquely identifying a structure while accounting for constraints,
such as translational
invariance; additionally, such a formulation is independent of the charge
density grid. 
% Using
%many-body correlation descriptors allows us to systematically improve the
%accuracy of the model by incorporating higher order correlations. Amplitudes
%of many-body descriptors used to predict the total energy allows us to uniquely
%identify a structure, account for constraints like translational invariance,
%and are not dependent on the charge density grid.
\end{abstract}

\maketitle

\section{Introduction}

Kohn-Sham density functional theory (DFT) is a popular technique that is
commonly used to calculate the electronic structure of a wide variety
of materials to predict, and/or to analyze their mechanical, optical,
electronic, or magnetic properties. This success can be attributed
to the fact that many physical and chemical properties of interest can be derived
from the ground state electronic density, $\rho \left( {\bf r}
\right)$. Here, $\rho \left( {\bf r} \right)$ is a scalar field and
${\bf r}$ corresponds to a point in the supercell. Typically, DFT
calculations are limited to supercells containing a few hundred
atoms, and due to the high computation cost associated with solving
the eigenvalue problem of the Kohn-Sham Hamiltonian, typical systems
in ab initio molecular dynamics (MD) simulations can be evolved 
only for a few tens of pico-seconds. This is orders of magnitude
smaller than time-scales associated with physical processes such
as phase transitions, conformational changes in molecules, or creep
failure of a material. Thus, it is desirable to develop surrogate
models that can speed up individual DFT calculations. Recently, a 
new set of methods based on machine learning approaches has been
proposed that can overcome or bypass the bottleneck of traditional
methods, and successfully predict molecular properties at lower
computational cost.\cite{KyleIssac2017} These techniques explore the physical information
of the data using statistical inference to train a model for the
desired properties. %learn the desired properties. 
The goal here is not to develop models that try to learn the Born-Oppenheimer
potential energy surface for the whole configuration space, but to
develop models that can learn only a part of the energy surface on-the-fly
and speed up MD or Monte Carlo simulations. For example, Brockherde
et al, in 2017 studied three different trained models:\cite{Brockherde2017}
$(i)$ the orbital-free map, $(ii)$ the Hohenberg-Kohn map, and
$(iii)$ the Kohn-Sham map. The orbital-free map method uses machine
learning to predict the kinetic energy functional for a given
ground state electronic density $\rho \left({\bf r} \right)$, and
uses its functional derivative to self-consistently solve the Euler
equation. On the other hand, the Kohn-Sham map attempts to model
the total energy as a functional of the external potential $v\left(
{\bf r} \right)$, and the Hohenberg-Kohn map is a two step procedure
that first attempts to learn $\rho \left({\bf r} \right)$ from $v
\left( {\bf r} \right)$ and then predict the total energy from $\rho
\left( {\bf r} \right)$. In their formulation, $v\left( {\bf r} \right)$
is obtained from the superposition of Gaussian functions placed at
the atomic positions and $\rho \left( {\bf r} \right)$ is represented
using Fourier basis functions. These methods can be used to model total
energies and charge density of small molecules like water,
benzene and ethane. But, the fact that hundreds of structures are
required to train models for systems containing only a few atoms poses
a serious practical problem in using these for large systems (for
example, crystalline solids, liquids). Thus, there is a need to develop
models that can be trained using a much smaller data set.

The Hohenberg-Kohn map has many advantages: it is computationally
more efficient than a direct DFT calculation and it is easy to parallelize
on large computers. In addition, representing the nuclear potential
using a set of Gaussians (weighted by their corresponding atomic numbers)
allows trained models to easily handle multicomponent systems
without increasing the computation cost. On the other hand, in many
methods that have been proposed to predict $\rho \left( {\bf r} \right)$
directly from a set of atomic positions, the number of features/descriptors
increases exponentially with the number of different elements
present in the system. Such a problem can perhaps be circumvented
by the method proposed by Ji and Jung in Ref [\onlinecite{JiJungJCP2018}]. These authors
proposed a local environment descriptor
%to model the electronic charge density
based on the conjecture that information about the
local environment surrounding a grid point can be compressed by
spherical averaging. Therefore, the authors generated feature vectors
for each grid point by averaging of pseudo-potentials over spheres of
a pre-defined set of radii. Motivated
by these results it is intuitive to ask if this feature vector can
be used to directly predict the total energy from an external
potential generated by a superposition of Gaussians (proposed by Brockherde
et al.)\cite{Brockherde2017} or a superposition of pseudo-potentials
(proposed by Ji and Jung)\cite{JiJungJCP2018}. 

Alternatively, Brockherde et al. proposed using a set of basis functions
to represent $\rho \left( {\bf r} \right)$ in order to decrease the
computational cost.\cite{Brockherde2017} Here we note that in addition to the computational
cost issue, for three-dimensional systems, such as crystalline solids or
liquids, simple constraints like global translation and rotation,
are difficult to implement in a grid-based representation framework.
Thus, many researchers have also modeled the electronic charge density
using a  set of basis functions. For example, Grisafi et al. modeled the
charge density using a atom-centered, symmetry-adapted Gaussian process regression
framework.\cite{GrisafiACS2019, GrisafiWilkinsCsyaniCeritti} Fabrizio
et al. in Ref [\onlinecite{C9SC02696G}] used a similar approach to predict
the ground state electronic charge density of a large number of dimers
by using specialized basis sets, such as Weigend's JK-fit cc-pVQZ. 

Many researchers have also used neural networks to generate
surrogate models. For example, Nagai et al.\cite{NagaiArXiv2019} %(arxiv 1903.00238)
modeled the electronic density 
and total energy with a flexible feed-forward neural network in
which the exchange-correlation potential is obtained by taking
functional derivatives using the back-propagation method. To
make the model independent of the grid, the authors used quantities
such as the local electronic density, local spin density and the scaled
gradient of the electronic density. %as features to model the
%total energy. %The success of this scheme in accurately
%reproducing functional derivatives means that the authors were
%able to overcome the hurdle faced by Brochard et al. in
%using a self-consistent framework. Many other researchers have
%used a set of features to train models. For example, 
% (npj computational materials, (2019) 5:22 ; https://doi.org/10.1038/s41524-019-0162-7)
Similarly, Chandrasekaran et al. used a variety of scalar, vector and tensor
% https://www.nature.com/articles/s41598-017-01251-z
fingerprints % and the charge density
to train a neural network model.\cite{Chandrasekaran2019}
On the other hand, in the PROPhet package developed by Kobl et al.,
the electronic charge density is predicted from the atomic position
by training a neural network model that uses descriptors proposed
by Behler and Parrinello.\cite{Kolb2017} 

% ALSO SEE: https://iopscience.iop.org/article/10.1088/2515-7639/ab0b4a/pdf
% VERY IMPORTANT: 1811.08928.pdf
Sinitskiy and Pande, in contrast, proposed a deep neural network based
scheme to capture subtle features of the electronic charge
density, such as lone pairs and hybridized electronic shapes around
aromatic and cyclopropane rings, with high accuracy.\cite{Sinitskiy2018}
Similarly, Dick and Fernandez-Serra have proposed a scheme %in which the
%electronic densities are used to train a neural network
to correct
baseline DFT energies and forces to the accuracy provided by a higher
level method.\cite{SebastianDick2019} From the computational point of
view, such methods are very promising because the time required for
charge density and total energy predictions are at least an order of
magnitude lower than brute force DFT calculations.
%However, it is not clear how such models can handle a three-dimensional
%system with periodic boundary conditions for which the model has to
%satisfy global rotation and translation constrains.
The advantages of such an approach, however, motivate the development
of a similar framework for three-dimensional systems (with periodic
boundary conditions) that satisfies global rotation and translation
constraints.
%Motivated by these
%developments, we ask if a similar framework can be developed for
%three-dimensional crystalline solids?
%In addition, even for small
%molecules the proposed methods require a large training data set
%which can computationally expensive.
In addition, the necessity of a large training data set even in the
cases where the preceding method is computationally viable serves as
an additional motivating factor for the development of a method which
is equally effective with less data. % Therefore, it is desirable
% to have methods that can learn over a smaller data set.

Eric Schmidt and co-workers in Ref [\onlinecite{SCHMIDT2018250}] modeled the ground state
electronic density using a linear model that specifically included 
two- and three-body correlations:
\begin{equation}
  \rho_{\rm e} \left( {\bf r} \right) = \sum_{\rm i = 1}^{N} \; f_{2} \left( \left|
      {\bf r} - {\bf r}_{\rm i}\right|\right) + \sum_{\rm i, j = 1}^{N} \;
      f_{3} \left( \left| {\bf r} - {\bf r}_{\rm i}, {\bf r} - {\bf r}_{\rm j}
      \right| \right).
      \label{EricModel1}
\end{equation}
Here $N$ is the total number of atoms, $i$ and $j$ are indices of atoms
located at ${\bf r}_{i}$ and ${\bf r}_{j}$, respectively, and $f_{2}$ and
$f_{3}$ are functions of one and two variables that capture two- and
three-body correlations. In Ref [\onlinecite{SCHMIDT2018250}], %these functions were modeled
%by using cosine basis functions.
it is reported that including the
three-body contributions dramatically improves the predictive capability of the
model as compared to a model containing only the two-body term. Next,
the total energy is calculated from a sum of two contributions: (a) an
embedding energy term that uses only the two-body contributions to the charge
density in Eq. $\ref{EricModel1}$, and (b) a pair energy term that depends
on the distance between an atom and its neighbors.
%term that the two-body correlation 
%to predict the total energy, the authors used the embedded atom
%method type functional form to capture the effect of the electronic
%density.
Since this model was tested on a wide variety of systems (including
metals, alloys, semi-metal, ceramic oxides), it is natural to ask
if it is possible to develop a generalized representation of the
electronic density that includes two-, three-, four-body and higher
order correlation. Similarly, it is interesting to explore if the
model to predict the total energy can also be systematically improved
by including three-body and higher order correlations. If this is possible, then
one can switch on or off terms based on the accuracy and computational
cost trade-offs.

Here, we propose a general framework to learn a model of the ground state charge
density using a many-body expansion. %The motivation is
We show that the
predictive capability can be enhanced by systematically incorporating
two-, three-, four-body or higher order correlations. The total energy
calculations, on the other hand, relies on a simple model that uses
amplitudes of the many body contributions (to the charge density) as
descriptors. These two models illustrate that these off-the-grid descriptors
are enough to infer the energy landscape at a relatively small cost.
In addition, a majority of the studies reported in the literature have
focused on isolated systems. In contrast, we focus on full three dimensional
periodic systems and explore avenues to predict $\rho \left( {\bf r}
\right)$ and the total energy using different nonlinear regression and
dimension reduction techniques.

The remainder of the paper takes the following form. In Section II, we
present details of our models. % to learn a model of the ground state electronic
%charge density and the total energy of the system.
In Section III,
we use these models to predict ground state charge densities and
total energies of amorphous structures of germanium. To this end,
we analyze the importance of higher order correlations in
accurately predicting the ground state charge density and compare
models obtained from a variety of linear and nonlinear regression techniques.

\section{Models to predict the electronic charge density and the total energy}

In the following discussion, in line with the Hohenberg-Kohn map
framework, we present two methods: one to predict the ground-state
electronic charge density from the atomic structure,
and another to predict the total energy from the ground state
electronic charge density. To predict $\rho \left( {\bf r} \right)$,
we propose an off-the-grid technique that relies on a set of descriptors
that can effectively capture the local environment around an atom.
As we show below, descriptors based on the two-, three-, four-body
correlations can effectively capture the underlying symmetry of the
distribution of atoms. These descriptors are then used to
train a linear regression based model using a set of training structures
and their corresponding ground state charge densities.

To predict the total energy from the electronic charge density, the
ground state electronic density of a given structure is first mapped
to a unique point in the feature space. This representation is
important to make the total energy predictions invariant to global
rotation/translation of $\rho \left( {\bf r} \right)$ (and atomic
positions) and to the ordering of grid indices. Next, we use a
variety of linear and nonlinear regression techniques to predict
the total energy from these features.

\subsection{Local correlation descriptors to predict the electronic density}

In this section, we propose a set of descriptors that can be
used to predict the ground state electronic charge density with
high accuracy for three dimensional systems using as few structures,
and corresponding charge densities, as possible for training.
Since, a few methods have already been proposed with this objective,
we want to develop a scheme that can allow us to systematically
improve the accuracy, not simply by increasing the number of basis
functions used as descriptors, but by adding higher order correlations
thereby allowing us to control the trade-off between accuracy and
computation cost. To design descriptors that embed local correlations,\cite{FerreJCP2015, GrisafiWilkinsCsyaniCeritti, Ralf2019prb, GrisafiACS2019, SamantaJCP2018, LinfengPRL2018, LingfengPRM2019, BartokRisiGabor2013, GlielmoPeterAlessandro2017, MatthiasAlexandreLilienfeld2012, ChimesNirGoldman2017, NirGoldman2016, TakahashiSekoTanaka2018, BartokGAP2010, BotuRamprasad2015, SekoTakahashiTanaka2014, Shapeev2016siam, Gastegger2018}
we first consider how the effective electronic density of an atom is
affected by the presence of other atoms. To this end, let us consider
a structure $S_{u}$ ($u$ is the structure identification index) at a
point ${\bf X}\in {\bf R}^{3N}$ in the configuration space containing
$N$ atoms, located at $\{ {\bf r}_{1}, {\bf r}_{2}, {\bf r}_{3}, \cdots,
{\bf r}_{N}\}$, such that ${\bf   r}_{i}\in {\bf R}^{3}$ for $i$ = 1,
2, ... , $N$. In addition, we assume the electronic density
of an atom, with index $i$, when it is isolated from any other atoms, is
represented by a smooth and differentiable function $\rho_{1} \left(
{\bf r}, {\bf r}_{i}\right)$. We 
represent the effective electronic density of this atom, when placed
amongst a distribution of other atoms (placed at ${\bf r}_{1}$,
${\bf r}_{2}$, $\cdots$, ${\bf r}_{Ni}$), by
\begin{equation}
  \bar{\rho} \left({\bf r}, {\bf r}_{i} | {\bf r}_{1}, {\bf r}_{2},
  \cdots, {\bf r}_{Ni} \right) = p_{1} - p_{2} + p_{3} - p_{4} + \cdots
  \label{manybody_density}
\end{equation}
Here, $p_{1} = \rho_{1} \left( {\bf r}, {\bf r}_{i}\right)$, $N_{i}$ is
the number of neighbors of $i$ and $p_{k}$ is the overlap between the
densities of the atom at ${\bf r}_{i}$ and its $k$ neighbors. Hence, $p_{k}$ depends
on the positions of neighbors of atom $i$ meaning that $\bar{\rho}
\left({\bf r}, {\bf r}_{i} | {\bf r}_{1}, {\bf r}_{2}, \cdots, {\bf r}_{Ni}
\right)$ can capture many-body effects. For example,
two-body correlations can be captured by the overlap between electronic
densities of atom $i$ and its neighbors

\begin{equation}
  p_{2} \left({\bf r}, {\bf r}_{i} | {\bf r}_{1}, {\bf r}_{2},
  \cdots, {\bf r}_{Ni} \right) = \sum_{j} \rho_{1}\left( {\bf r},
        {\bf r}_{i}\right) \rho_{1}\left( {\bf r}, {\bf r}_{j}\right) .
  \label{twobodycorrelation}
\end{equation}
Here, the sum is over all possible neighbors of atom $i$ in the
system. Similarly, the overlap between the electronic densities
of atom $i$ and two of its neighbors (with indices $j$ and $k$)
is given by (see Fig. $\ref{Correlations3body}$)
\begin{equation}
  \begin{split}
    &p_{3} \left({\bf r}, {\bf r}_{i} | {\bf r}_{1}, {\bf r}_{2},
    \cdots, {\bf r}_{Ni} \right)\\
    &= \sum_{j, k} \rho_{1}\left( {\bf r}, {\bf r}_{i}\right) 
    \rho_{1}\left( {\bf r}, {\bf r}_{j}\right)  
    \rho_{1}\left( {\bf r}, {\bf r}_{k}\right).
  \end{split}
  \label{threebodycorrelation}
\end{equation}

Following the same procedure as mentioned above, we can design
descriptors to capture higher order correlations. For example,
the overlap between electronic densities of atom $i$ and three 
of its neighbors is given by
\begin{equation}
  \begin{split}
    &p_{4} \left({\bf r}, {\bf r}_{i} | {\bf r}_{1}, {\bf r}_{2},
    \cdots, {\bf r}_{Ni} \right) \\
    &= \sum_{j, k, l}  
    \rho_{1}\left( {\bf r}, {\bf r}_{i}\right) 
    \rho_{1}\left( {\bf r}, {\bf r}_{j}\right)  
    \rho_{1}\left( {\bf r}, {\bf r}_{k}\right)
    \rho_{1}\left( {\bf r}, {\bf r}_{l}\right).
    \end{split}
    \label{fourbodycorrelation}
\end{equation}
Here $j$, $k$ and $l$ are indices of three atoms in the neighborhood 
of atom $i$. Thus, if $\rho \left( {\bf r}, {\bf r}_{i} \right)$ is
known, the spatial variation in the effective density in
Eq. $\ref{manybody_density}$ can be calculated by including the different
correlations. Here we note that Eqs.
$\ref{twobodycorrelation}$-$\ref{fourbodycorrelation}$ describe the
correlation between atoms, and ${\bf r}$ can be the spatial location
of a grid point. Thus, in the following sub-sections, we use this
notion to propose descriptors that can very effectively capture the
local neighborhood information around a grid point.

\subsubsection{Two-body correlations} For a systematic analysis of
correlations, the functional form of the electronic density
$\rho_{1} \left( {\bf r}, {\bf r}_{j} \right)$ of an isolated atom
(located at ${\bf r}_{j}$) is required. This quantity can be calculated
using any first principles methodology, but to keep the formulation
general, we use a set of orthogonal basis functions to capture
these correlation descriptors.

To calculate two-body correlations, we need to capture the effect of
the electronic density of an atom (located at ${\bf r}_{j}$) on the
grid point (located at ${\bf r}$). Thus, we define
\begin{equation}
  \begin{split}
    &\sum_{j} \rho_{1} \left( {\bf r}, {\bf r}_{j} \right) = \sum_{n}
    c_{n}^{\left( 21 \right)} \phi_{n}^{\left( 21 \right)}, \\
    &{\rm where}, \;\; \phi_{n}^{ \left( 21 \right)} =
    \sum_{j} \psi_{n} \left( \left| {\bf r} - {\bf r}_{j} \right| \right).
  \end{split}
  \label{twobodycorrelationPsi}
\end{equation}
Here, $\psi_{n}$ ($n$ = 1, 2, 3, ...) is a set of orthogonal basis
functions, ${\bf r}$ is the spatial location of the grid point, the
atom with index $j$ is in the neighborhood of this grid point, and
$c_{n}^{\left( 21\right)}$ are the coefficients of this expansion.
It is worth noting that this two-body correlation captures the proximity
between a grid point and an atom. Hence, it is different from the
two-body correlation between two atoms that is defined in Eq. $\ref{twobodycorrelation}$

\subsubsection{Nomenclature} To describe the different types of many-body
correlations we use the following nomenclature. The number $p$ in
$c^{ \left( pq\right)}$ is the total number of atoms and grid points
present in the correlation and $q$ denotes the number of bonds. For example,
in a two-body correlation, an atom and a grid point are connected by a
single bond. Thus, $q = 1$ and $p = 2$.
%there is an atom and a grid point in the two-body correlation and there is
%only a single bond between the grid point and an atom in its neighborhood,
%so $q = 1$ and $p = 2$.
Similarly, $c^{ \left( 33 \right)}$ and $c^{ \left( 32\right)}$ correspond
to three-body correlations with three and two bonds, respectively (see
Fig. $\ref{Correlations3body}$).
%corresprepresents
%coefficients that capture three-body correlation containing three bonds
%(see Fig. $\ref{Correlations3body}$) and $c^{ \left( 32\right)}$ corresponds
%to three-body correlations with two bonds.
The number of bonds (i.e.
$q$) present in a correlation determines the number of orthogonal basis
functions being multiplied and the computation cost increases exponentially
as both $p$ and $q$ increases.

\subsubsection{Three-body correlations}
Next, we consider descriptors based on three-body correlations. Two important
three-body correlations that arise when we consider a grid point and 
%from the overlap 
%between electronic densities of an atom with index $i$ and
two atoms, with indices $j$ and $k$, in the neighborhood are detailed below
(see Fig. $\ref{Correlations3body}$).
\begin{itemize}
\item[i.] The three-body correlation that arises
  % we need to account for this is
  due to the overlap between the electronic densities of atoms
  ${j}$ and ${k}$ is:
  \begin{equation}
    \begin{split}
      \rho^{\left( 32\right)} &= \sum_{j, k} \rho_{1} \left( {\bf r}, {\bf r}_{j}\right) 
      \rho_{1} \left( {\bf r}, {\bf r}_{k}\right) = \sum_{m, n} c_{mn}^{\left( 32\right)} 
      \; \phi_{mn}^{\left( 32\right)}, \\
      \phi_{mn}^{\left( 32\right)} &= \sum_{j, k} \psi_{m}
      \left( \left| {\bf r} - {\bf r}_{j} \right| \right)
      \psi_{n} \left( \left| {\bf r} - {\bf r}_{k} \right| \right).
    \end{split}
    \label{c32correlation}
  \end{equation}
  Here, $c_{mn}^{\left( 32\right)}$ are the coefficients of this expansion.
  It is easy to see that $\rho^{\left( 32\right)}$ is not sensitive to the
  distance between atoms $j$ and $k$, i.e. it does not change if atoms $j$
  and $k$ are located anywhere on the surface of a sphere centered at
  ${\bf r}$.
  
\item[ii.] Next, we consider a correlation that also encodes the proximity
  between the atoms $j$ and $k$:
  \begin{equation}
    \begin{split}
      &\rho^{\left( 33\right)} 
      = \sum_{m, n, o} c_{mno}^{\left( 33\right)} \; \phi_{mno}^{\left( 33\right)},
      \;\;\; {\rm where}\\
      &\phi_{mno}^{\left( 33\right)} =
      \sum_{j, k} \psi_{m}
      \left( \left| {\bf r} - {\bf r}_{j} \right| \right)
      \psi_{n} \left( \left| {\bf r} - {\bf r}_{k} \right| \right)
      \psi_{o} \left( \left| {\bf r}_{j} - {\bf r}_{k} \right| \right).
    \end{split}
    \label{c33correlation}
  \end{equation}

  Schematic illustrations of these correlations are shown %illustrated graphically
  in Fig. $\ref{Correlations3body}$. 
  
  \begin{figure}[htp]
    \centering
    \includegraphics[width=0.35\textheight]{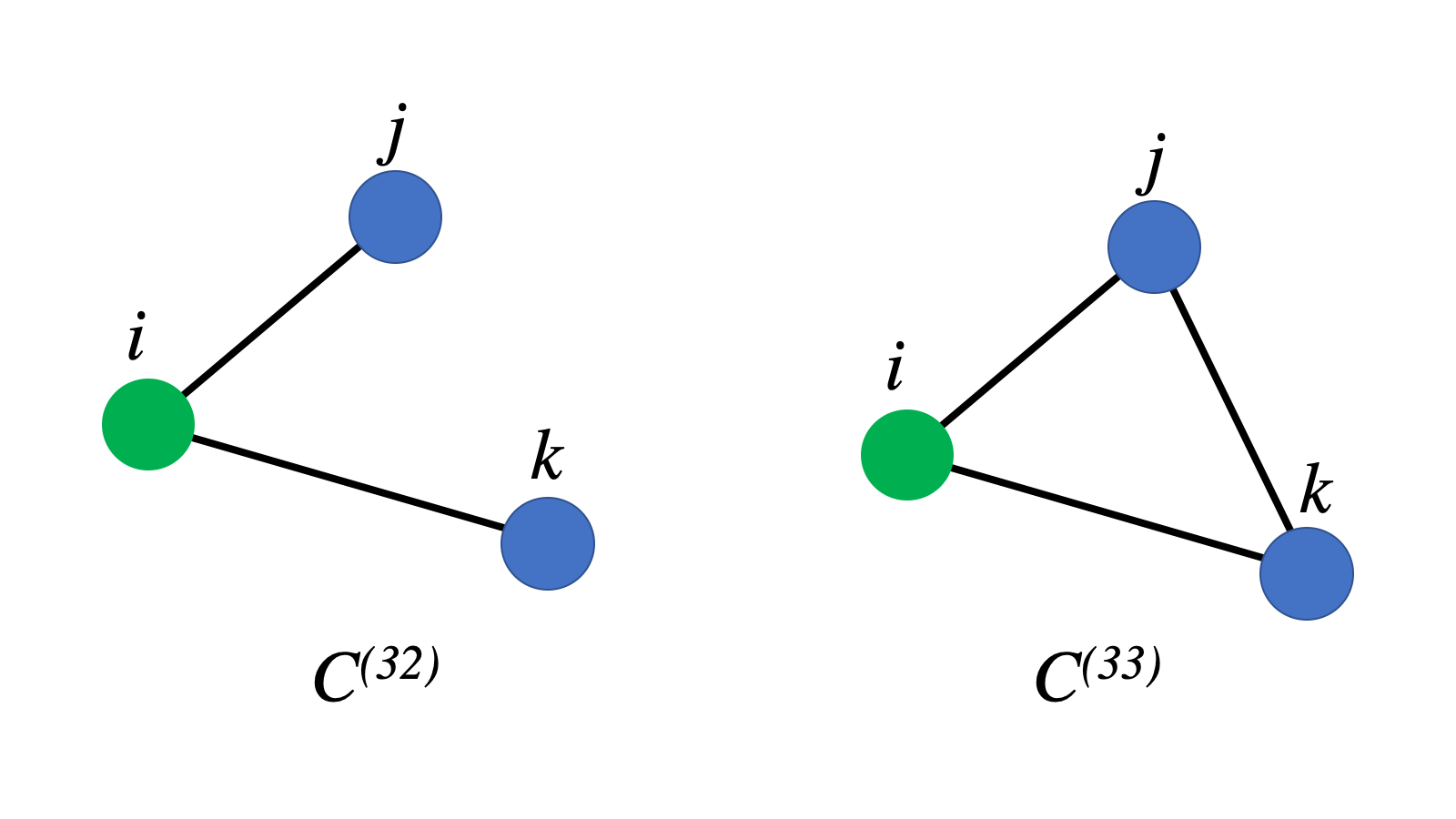}
    \caption{A schematic representation of correlations due to the overlap
      between densities of three atoms. Each bond represents an overlap
      between electronic densities of two atoms. Closed paths or loops
      represent overlap between densities of all three atoms.}
    \label{Correlations3body}
  \end{figure}
  
\end{itemize}

\subsubsection{Four-body correlations}
\label{subsec_Corr4body}
 
To obtain descriptors based on four-body correlations, we consider a grid
point and overlaps between the electronic densities of three atoms with
indices ${j}$, $k$ and $l$. These contributions are detailed below:

\begin{itemize}
\item[iii.] The correlation arising due to the overlap between the
  electronic densities of three atoms ${j}$, ${k}$ and ${l}$
  at the grid point is captured by (see Fig. $\ref{Correlation43}$)
  \begin{equation}
    \begin{split}
      &\rho^{\left( 43 \right)} = \sum_{j, k, l} 
      \rho_{1}\left( {\bf r}, {\bf r}_{j}\right) 
      \rho_{1}\left( {\bf r}, {\bf r}_{k}\right) 
      \rho_{1}\left( {\bf r}, {\bf r}_{l}\right) \\
      &= \sum_{m, n, o}
      c_{mno}^{\left( 43\right)} \; \phi_{mno}^{\left( 43\right)}, \;\; \; {\rm where},\\
      &\phi_{mno}^{\left( 43\right)} = 
      \sum_{j, k, l} \psi_{m} \left( \left| {\bf r} - {\bf r}_{j} \right| \right)
      \psi_{n} \left( \left| {\bf r} - {\bf r}_{k} \right| \right)
      \psi_{o} \left( \left| {\bf r} - {\bf r}_{l} \right| \right).
    \end{split}
    \label{c43correlation}
  \end{equation}

  \begin{figure}[htp]
    \centering
    \subfigure[] {
      \includegraphics[width=0.35\textheight]{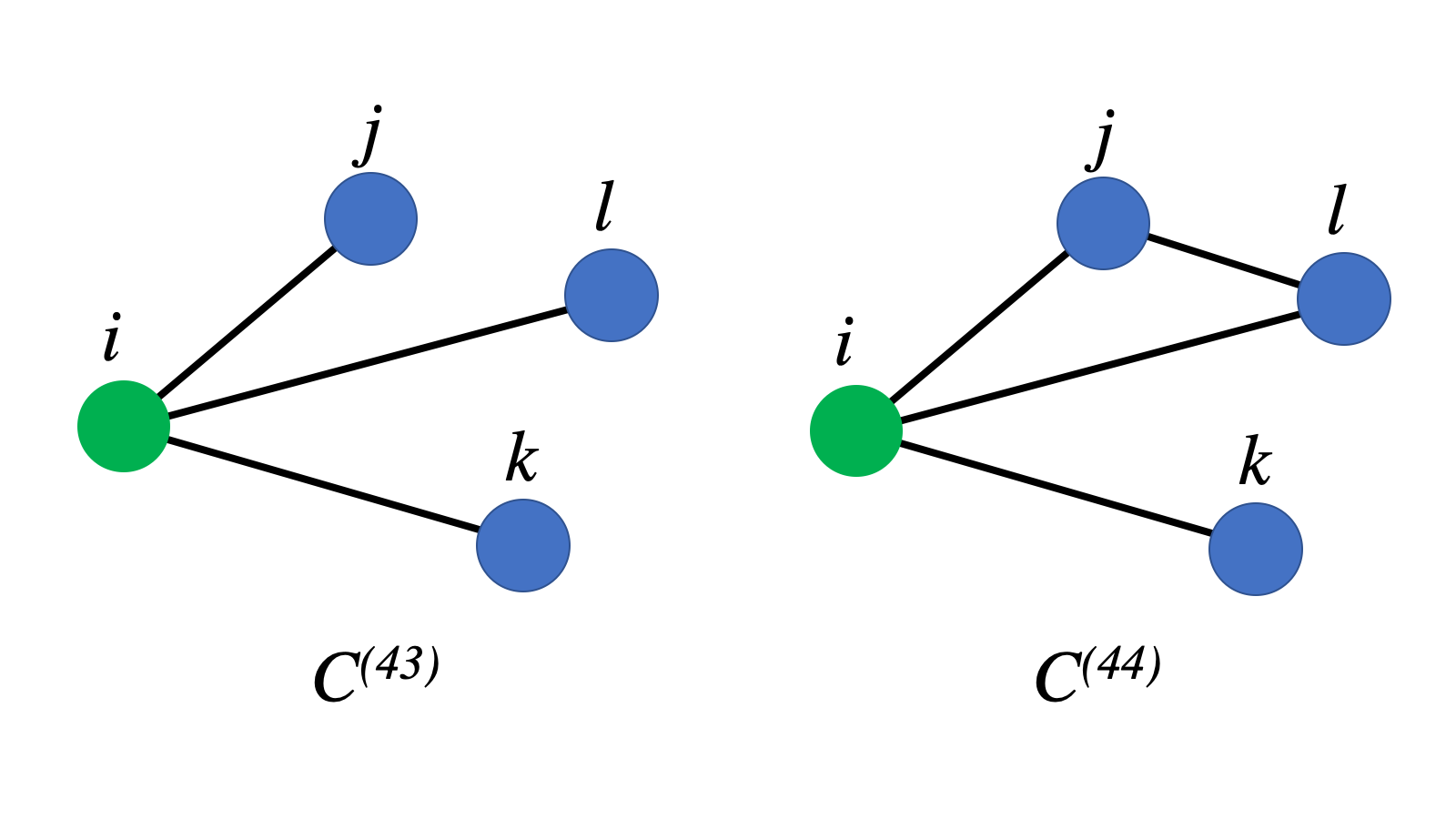}
      \label{Correlation41}
    }
    \subfigure[] {
      \includegraphics[width=0.35\textheight]{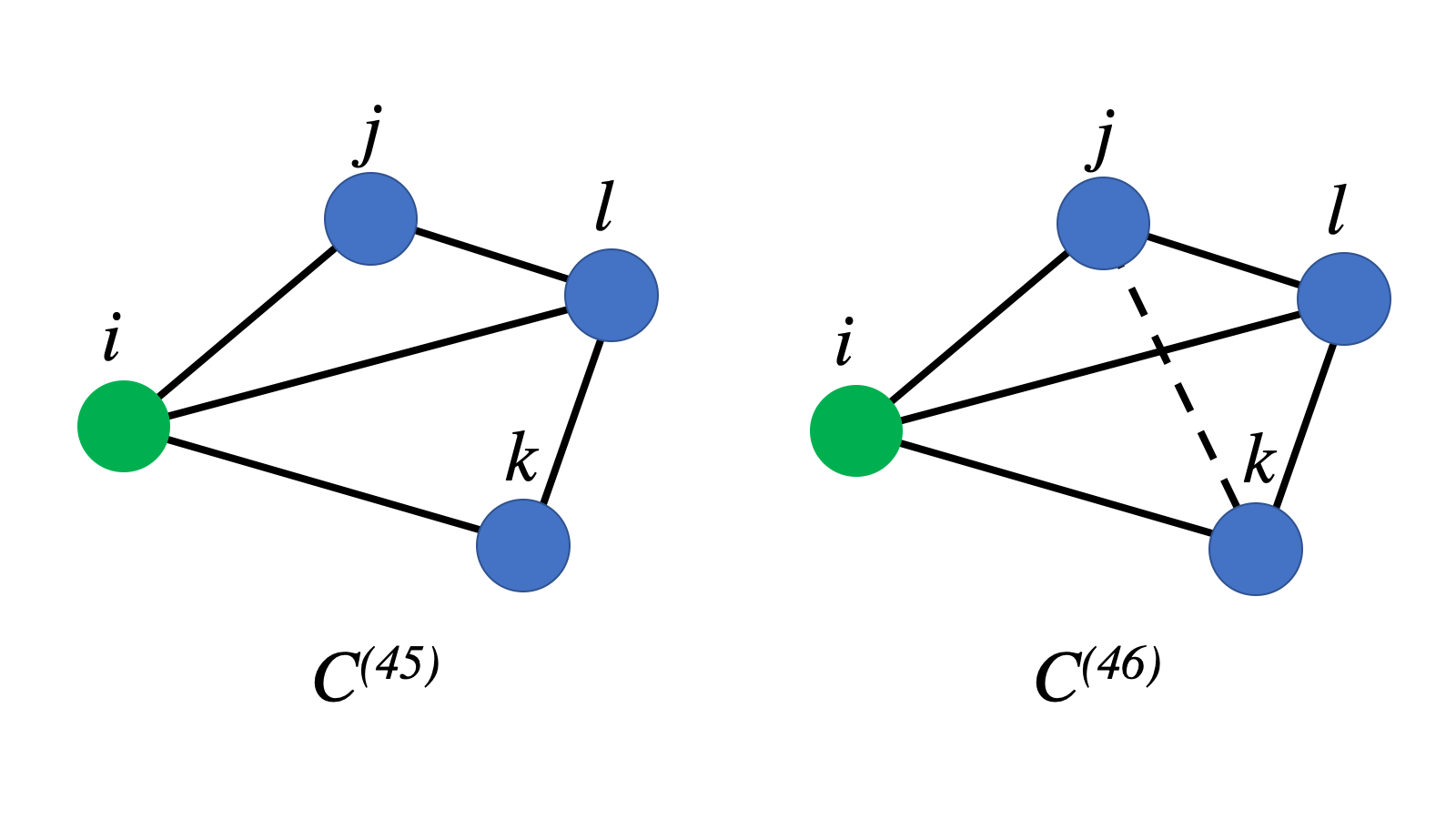}
      \label{Correlation43}
    }
    \caption{Correlations due to the overlap between the densities in
      a cluster containing 4 atoms.}
    \label{Correlations4body}
  \end{figure}
  
\item[iv.] The correlation that encodes the separation between a pair of
  neighbors of the grid point is given by
  \begin{equation}
    \begin{split}
      &\rho^{\left( 44 \right)} = \sum_{m, n, o, p}
      c_{mnop}^{\left( 44\right)} \; \phi_{mnop}^{\left( 44\right)}, \;\;\; {\rm where}\\
      &\phi_{mnop}^{\left( 44\right)} =
      \sum_{j, k, l} \psi_{m} \left( \left| {\bf r} - {\bf r}_{j} \right| \right)
      \psi_{n} \left( \left| {\bf r} - {\bf r}_{k} \right| \right) \\
      &\qquad \qquad \qquad \times\psi_{o} \left( \left| {\bf r} - {\bf r}_{l} \right| \right)
      \psi_{p} \left( \left| {\bf r}_{j} - {\bf r}_{l} \right| \right)
    \end{split}
    \label{c44correlation}
  \end{equation}

  Geometrically, this correlation accounts for an extra edge 
  between the two neighbors $j$ and $l$ of the grid point as shown 
  in Fig. $\ref{Correlation41}$. 

\item[v.] Next, we consider the correlation that arises when two pairs
  of atoms are connected by edges. For example,
  in Fig. $\ref{Correlation43}$, two pairs of neighbors ($j$, $l$) and
  ($k$, $l$) are connected by extra edges and the corresponding four
  body correlation is given by
  \begin{equation}
    \begin{split}
      &\rho^{\left( 45 \right)} = \sum_{m, n, o, p, q} c_{mnopq}^{\left( 45 \right)} \;
      \phi_{mnopq}^{\left( 45\right)}, \;\;\; {\rm where}\\
      &\phi_{mnopq}^{\left( 45\right)} =
      \sum_{j, k, l} \psi_{m} \left( \left| {\bf r} - {\bf r}_{j} \right| \right)
      \psi_{n} \left( \left| {\bf r} - {\bf r}_{k} \right| \right) \\
      &\qquad \; \times\psi_{o} \left( \left| {\bf r} - {\bf r}_{l} \right| \right)
       \psi_{p} \left( \left| {\bf r}_{j} - {\bf r}_{l} \right| \right)
       \psi_{q} \left( \left| {\bf r}_{k} - {\bf r}_{l} \right| \right)
    \end{split}
    \label{c45correlation}
  \end{equation}

\item[vi.] Finally, we consider the correlation that arises when
  all the three neighbors of a grid point are connected by edges.
  \begin{equation}
    \begin{split}
      &\rho^{\left( 46 \right)} = \sum_{m, n, o, p, q, r}
      c_{mnopqr}^{\left( 46\right)} \; \phi_{mnopqr}^{\left( 46\right)}, \;\;\; {\rm where},\\
      &\phi_{mnopqr}^{\left( 46\right)} = 
      \sum_{j, k, l} \psi_{m} \left( \left| {\bf r} - {\bf r}_{j} \right| \right)
      \psi_{n} \left( \left| {\bf r} - {\bf r}_{k} \right| \right) 
      \psi_{o} \left( \left| {\bf r} - {\bf r}_{l} \right| \right) \\
      &\qquad \; \times \psi_{p} \left( \left| {\bf r}_{j} - {\bf r}_{l} \right| \right)
       \psi_{q} \left( \left| {\bf r}_{k} - {\bf r}_{l} \right| \right)
       \psi_{r} \left( \left| {\bf r}_{k} - {\bf r}_{j} \right| \right)
    \end{split}
    \label{c46correlation}
  \end{equation}
 
\end{itemize}

Intuitively, this procedure can be repeated to generate higher order descriptors.
But, as we show in the subsequent sections, our analysis suggests that
uncertainties in models containing up to four-body correlations are very low.

\subsubsection{Linear regression model to predict $\rho \left( {\bf r} \right)$}

Using these different two, three, four-body correlations we
obtain the following model to represent the electronic density at a grid
point (with indices ($i$, $j$, $k$) and spatially located at ${\bf r}_{ijk}$)
\begin{equation}
  \begin{split}
    &\rho \left( {\bf r}_{ijk} \right) = \sum_{n} c^{
      \left( 21\right) }_{n} \phi^{ \left( 21\right) }_{n} +
    \sum_{m, n} c^{ \left( 32 \right) }_{mn} \phi^{ \left( 32\right) }_{mn} + \\
    &\sum_{m, n, o} \left[ c_{mno}^{\left( 33 \right)} \phi_{mno}^{\left( 33\right)} + 
      c_{mno}^{\left( 43\right)} \phi_{mno}^{\left( 43\right)} \right] + \\
    &\sum_{m, n, o, p} \left[ c_{mnop}^{\left( 44 \right)} \phi_{mnop}^{\left( 44\right)} + 
      c_{mnop}^{\left( 54 \right)} \phi_{mnop}^{\left( 54\right)} \right] + \cdots
    \label{ElectronicDensityLinearModel1}
  \end{split}
\end{equation}

Let the target electronic density at all the grid points %used for training
% the model
be represented by the vector $\boldsymbol{\rho}$ and ${\bf c}$ be
a vector that contains all the correlation coefficients, i.e. $\{c^{ \left(
  21\right) }_{1}, c^{ \left( 21\right) }_{2}, \cdots , c^{ \left( 32 \right)
}_{11}, c^{ \left( 32 \right) }_{21}, \cdots \}$. Then,
Eq. $\ref{ElectronicDensityLinearModel1}$ can be expressed as a linear
system of equations as shown below
\begin{equation}
  \boldsymbol{\rho} = \mathbb{M} {\bf c}.
  \label{ElectronicDensityLinearModel2}
\end{equation}
Here, $\mathbb{M}$ is a matrix, such that the number of rows is equal to
the total number of grid points in the training data set, and the number
of columns is equal to the total number of correlations present in the
many-body expansion in Eq. $\ref{ElectronicDensityLinearModel1}$. To
obtain the coefficients ${\bf c}$ of this many-body expansion, we use a
linear regression framework \cite{elements,bishop} and
minimize the following cost function with $L_{2}$-regularization:
\begin{equation}
  \Omega_{1} = \sum_{S_{u}} \sum_{i, j, k} \; \left|\rho_{ijk} \left({\bf c} \right) - \rho_{ijk}^{\rm DFT}
  \right|^{2} + \lambda_{1} \left| {\bf c} \right|^{2}_{L_{2}}.
  \label{eq:linden}
\end{equation}
Here $\rho_{ijk} \left({\bf c} \right) = \rho \left( {\bf r}_{ijk} \right)$
is the density described in Eq. $\ref{ElectronicDensityLinearModel1}$
at grid point ($i, j, k$), % of a given charge density in the training
% data set,
$\rho_{ijk}^{\rm DFT}$ is the charge density at the same
grid point obtained from a direct DFT calculation. The sum over $S_{u}$
is over all the structures and the corresponding charge densities in the
training data set. %Then the matrix $\mathbb{M}$ can
%be constructed from an arbitrary set of atoms close to a grid point
%and the density predicted using the set or vector of the
%coefficients of correlation amplitudes ${\bf c}$ obtained from the
%training data.
The model in Eq. $\ref{eq:linden}$ can be easily modified to
incorporate model selection using $L_{1}$-regularization, or
a mixture of $L_{1}$ and $L_{2}$-regularization.

\subsection{Total energy prediction}

In this section, we present a framework to predict the total energy of a
structure. Here we note that the total energy is invariant to global
rotation and global translation of the structure and the corresponding
ground state charge density. In addition, the total energy, $E_{u} =
E_{u} \left( \rho \left( {\bf r}_{u} \right) \right) $, of a structure
(denoted by index $u$) has to be independent of how grid points of the
charge density are indexed (i.e. permutation of grid indices). To mitigate
these issues, we propose to minimize the following cost function for each
structure present in the training data set:
\begin{equation}
  \bar{\Omega}_{1} = \sum_{i, j, k} \; \left|\rho_{ijk} \left(\bar{\bf
    c}^{\left( u\right)} \right) - \rho_{ijk}^{\rm DFT} \right|^{2} +
  \lambda_{1} \left| \bar{\bf c}^{\left( u \right)} \right|^{2}_{L_{2}}.
  \label{eq:linden1}
\end{equation}
Thus, $\Omega$ in Eq. $\ref{eq:linden}$ differs from $\bar{\Omega}_{1}$
in Eq. $\ref{eq:linden1}$ because of the additional summation (in
Eq. $\ref{eq:linden}$) over all structures in the training set. After
minimizing $\bar{\Omega}_{1}$, the ground state electronic density $\rho \left(
{\bf r}_{u} \right)$ for a structure (denoted by an index $u$) can be
represented by a unique coefficient vector which is denoted by $\bar{\bf
  c}^{\left( u \right)}$. Here we note
that the linear regression model proposed in Section IIA is to predict
the ground state charge density at each grid point based on its local
environment. % and is based on training a model for all structures in the
%training data set.
However, to predict $E_{u} \left( \rho \left( {\bf r}_{u} \right) \right)$
from $\rho \left( {\bf r}_{u} \right)$, we seek a set of coefficients
$\bar{\bf c}^{\left( u \right)}$ that can uniquely identify
the whole structure. %(and its ground state charge density $\rho \left(
%{\bf r}_{u} \right)$).
Now, to predict the total energy, we use these coefficients as descriptors
and train a variety of linear and nonlinear models using different regression techniques.

\subsubsection{Linear regression for total energy prediction}
Given a set of correlation coefficients (i.e. $\bar{\bf c}^{\left(1\right)}$,
$\bar{\bf c}^{\left(2\right)}$, $\bar{\bf c}^{\left(3\right)}$, $\cdots$ for
different structures in the data set), the linear regression model to predict
the total energy is given by
\begin{equation}
  \Omega_{2} = \left| E \left( \mathbb{C} \right) - E_{\rm DFT}
  \right|^{2} + \lambda_{2} \left| {\boldsymbol{\beta}} \right|^{2}_{L_{2}}, \;{\rm and},\; 
  E \left( \mathbb{C} \right) = \mathbb{C} \boldsymbol{\beta}
  \label{eq:linener}
\end{equation}
Here $\mathbb{C}$ is a matrix such that the $u$-th row of $\mathbb{C}$
is the coefficient vector $\bar{\bf c}^{\left( u\right)}$ of the $u$-th
configuration (as described in the previous paragraph) in the training
data set, and the number of rows is the number of configurations. The
coefficients of this linear model are represented by the vector $\boldsymbol{\beta}$.
%{\color{blue}The regularization coefficient $\lambda_{2}$ is an hyper-parameter that
%is optimized the bias-variance trade-off and improve the error on the testing set improving
%the performance on the unseen data.}

\subsubsection{Principal Component Regressions for the total energy prediction}

To generate a computationally viable model, we seek to minimize the
number of structures that are required to train the model in
Eq. $\ref{eq:linener}$. This means that the number of coefficients or
features for each configuration has to be smaller than the number of
samples. To solve this problem, we use the principal components
analysis (PCA) which is an unsupervised learning method that helps us
to build a low dimensional ($<$ dim$\left( {\boldsymbol{\beta}} \right)$)
representation of the coefficients $\bar{\bf c}^{\left( s\right)}$.

The central ideal of PCA is to minimize the error, denoted by $\epsilon$,
between the original data ${\bf X}$ and its low-dimensional representation
(which is denoted by $\hat{\bf X}$):
\begin{equation}
  \epsilon = \| \mathbf{X}-\mathbf{\hat{X}} \|_{\rm min}.
  \label{PCAerror}
\end{equation}
The PCA method uses a linear transformation to project the original data
to a subspace of maximum variance and an orthogonality constraint is 
imposed on this transformation. When the $\epsilon$ in Eq. $\ref{PCAerror}$
is minimized with this orthogonality constraint, we obtain the following
covariance matrix, $\Sigma$
\begin{equation}
  \Sigma = \frac{1}{m-1} \mathbf{X}^T\mathbf{X}.
  \label{PCAcovmat}
\end{equation}
Here, $m$ is the number of observations. 
Next, we define a matrix ${\bf U}$, that contains the eigenvectors of the
covariance matrix ordered in descending order of the magnitude of their
variance -- the first column corresponds to the eigenvector of maximum
variance and the last column corresponds to the eigenvector with the smallest
variance. Next, we select a set of $k$ eigenvalues (and their corresponding
eigenvectors) that account for most of the total variance (which is usually
more than 90\%). The eigenvectors corresponding to these $k$ selected
eigenvalues (sorted in the descending order) are stored in the column matrix
${\bf U} \left( k \right)$. The original data is then transformed to a low
dimensional representation by using 
\begin{equation}
  \hat{\bf X} = {\bf X} {\bf U} \left( k \right).
\end{equation}
In our case $\mathbf{X}=\mathbb{C}$, and we want to obtain a matrix
$\hat{\bf C}$ with smaller number of features, i.e. the dimensions of
the feature space $\hat{\bf c}$ for each configuration is smaller
 than $\bar{\bf c}$. Next, Eq. \ref{eq:linener} is solved using
 $\hat{\bf C}$ instead of $\mathbb{C}$.
   
 To evaluate the number of principal components necessary to get a good
 model, we analyze the convergence of the coefficient of determination,
 $R^2$, and the root mean square error, RMSE, with respect to the number
 of principal components. %The coefficient of determination
%$R^{2}$ is expressed in terms of the percentage variation of the data
%captured by the model:
%\begin{equation}
%  R^{2} = 1 - \frac{MSE}{Var}.
%\end{equation}
%Here $MSE$ and $Var$ stand for the mean squared error of the prediction and
%the variance of the observations, respectively.
A good fit corresponds to the case when $R^{2} \to 1$, i.e. the mean squared
error is much smaller than the variance of the data.

\subsubsection{Nonlinear regressions of the total energy}
\label{kernel-methods}

To capture existing nonlinearities in the representation of the total
energy using the features $\hat{ \bf c}$ (or $\bar{\bf c}$), we use
non-parametric regressions, such as kernel ridge and Gaussian process
regressions.\cite{elements, bishop, rupp} The kernel ridge regression
incorporates nonlinearity
into the model by using a nonlinear function also called the kernel
(denoted by $k \left( \bar{\bf c}_{i}, \bar{\bf c}_{j} \right)$ for 
two different structures $i$ and $j$) that captures the similarity
between two structures. The total energy of a test structure with
features ${\bf c}^{\ast}$ is then given by
\begin{equation}
  E^{*}= \sum_{i} \alpha_i \cdot k \left( \bar{\bf c}_{i},
  \bar{\bf c}^{\ast} \right)
  \label{kernelRR_energy}
\end{equation}
and the coefficients $\alpha$ are obtained by minimizing the cost
function in Eq. $\ref{eq:linener}$. The kernel function is positive
definite and it shapes the way the features are
compared in high dimensional space; the explicit equations to find
the coefficients of the kernel regressions can be found
elsewhere.\cite{rupp, bishop, mackay}

Gaussian process regression is a nonlinear regression technique
that is very similar to kernel ridge regression; the difference
arises from the assumption that the coefficients $\alpha_{i}$ in
Eq. $\ref{kernelRR_energy}$ are normally distributed in case of
the Gaussian process regression. As a consequence,
the regression output is also normally distributed.\cite{mackay} The
covariance of the output distribution and its error is defined by the
corresponding kernel, ${\bf K}$ which is given by
\begin{equation}
  {\bf K} = \bar{\bf K} + \sigma^{2} {\bf I}.
  \label{GP_bigKernel}
\end{equation}
Here $\bar{\bf K}_{ij} = k \left( \bar{\bf c}_{i}, \bar{\bf c}_{j}
\right)$ and $\sigma$ defines the strength of the noise, and its role is
similar to the regularization coefficient used in linear regression.
To model the total energies using the Gaussian process regression,
we use the radial basis function kernel, i.e. 
\begin{equation}
  K_{ij} = \lambda^{2} \exp \left( -\gamma^{2} \left| {\bf c}_{i} -
  {\bf c}_{j} \right|^{2} \right) + \sigma^{2} \delta_{ij}.
  \label{GP_bigKernel_RBF}
\end{equation}
This kernel has been successfully used for a wide range of
problems.\cite{mackay} In Eq. $\ref{GP_bigKernel_RBF}$, the parameter
$\gamma$ scales the difference between the feature vectors, $\lambda$
scales the exponential term, and $\sigma$ is the amplitude of the noise
and controls the strength of the regularization.   

The kernel and Gaussian process regression methods differ in the way
the hyper-parameters of the models are optimized. In the case of kernel
ridge regression, the commonly used model selection methods are the
leave-out-one, three way hold-out, or $k$-fold cross-validation. In
Gaussian process regression, the hyper-parameters are typically calculated
by either maximizing the
log-likelihood function, by using cross-validation along with the
log-likelihood method, or by using the Markov chain Monte Carlo method.
For the analysis presented here we use the three-way hold-out method
for cross-validation, and log-likelihood maximization to obtain the
model hyper-parameters for kernel ridge regression and Gaussian process
regression, respectively. 

\section{Results: Ground state charge density}

\subsection{DFT data for testing}
\label{data_set_description}

To train and test our model for the ground state charge density and
total energy predictions we use amorphous Ge as the model system.
This choice is motivated by the fact that many machine learning
techniques for interatomic potential generation fail to appropriately
capture the interactions between atoms in a liquid or in disordered
systems. For training and testing purposes, we generated two data sets:
\begin{enumerate}
\def\labelenumi{\alph{enumi})}
% \tightlist
\item
  {\it Data set A:} To obtain the first data set, we start with a
  disordered Ge supercell containing 512 atoms with edge lengths of
  23.5$\times 23.5\times 23.5$ ${\rm\AA}^{3}$. This structure is
  obtained by melting a crystalline solid and quenching the system
  to 100 K. From this reference structure, we generate 500 configurations
  by perturbing the atomic positions in the disordered Ge
  supercell by using random numbers uniformly distributed in the
  interval \((-\delta, \delta)\) $\rm\AA$, where $\delta$
  is set to 2\% of the supercell edge length.
\end{enumerate}

\begin{enumerate}
\def\labelenumi{(\alph{enumi})}
\setcounter{enumi}{1}
% \tightlist
\item
  {\it Data set B:} Next, we generate a second reference structure
  by perturbing (perturbation amplitude is 1\% of the edge length)
  the atomic positions of a disordered Ge structure. Using this reference
  structure as the seed, we generate another set of 500 perturbed
  structures by using random numbers uniformly distributed in the
  interval of (-1, 1) \(\rm{\mathring{A}}\).
\end{enumerate}

All density functional theory calculations are performed using the
plane wave basis density functional theory (DFT) implementation in
Vienna Ab-initio Simulation Package (VASP)$\cite{KresseF96, KresseFJ96}$
using the PBE$\cite{PerdewBE96}$ exchange correlation functional.
We use a plane wave cut-off of 560 eV to expand the wavefunctions and
a 2$\times$2$\times$2 ${\bf k}$-point mesh to perform the Brillouin
zone integrations. We use supercells containing 512 Ge atoms
and the supercell edge lengths are equal to 23.5 $\rm\AA$.

\subsection{Predicting the charge density}

To train a model that can accurately reproduce the ground state
charge density at each grid point, correlation functions described
in Section II are evaluated using a set of Chebyshev polynomials
of first kind. Chebyshev polynomials satisfy the following orthogonality
relation ($m$, $n > 0$)
\begin{equation}
  \int_{-1}^{1} \psi_{n} \left( s \right) \psi_{m} \left(s \right)
  \frac{ds}{\sqrt{1 - s^{2}}} = \frac{\pi}{2}\delta_{mn}
\end{equation}
To evaluate a Chebyshev polynomial, the input argument has to be
in the range of $[-1, 1]$. Thus, the distance, $r$, between an atom
and a grid point is scaled according to the following prescription
\begin{equation}
  s = \cos \left[ \frac{r - r_{\rm min}} {r_{\rm max} - r_{\rm min}} \right].
\end{equation}
Here, $r_{\rm max}$ and $r_{\rm min}$ are parameters that correspond
to the maximum and minimum distances between an atom and a grid point.
For computational efficiency purposes while evaluating the neighborhood
of a grid point, we introduce a cutoff radius. Thus, $r_{\rm max}$
is set to be equal the cutoff radius. 

\begin{figure*}[htp]
  \centering
  \subfigure[] {
    \includegraphics[width=0.23\textheight]{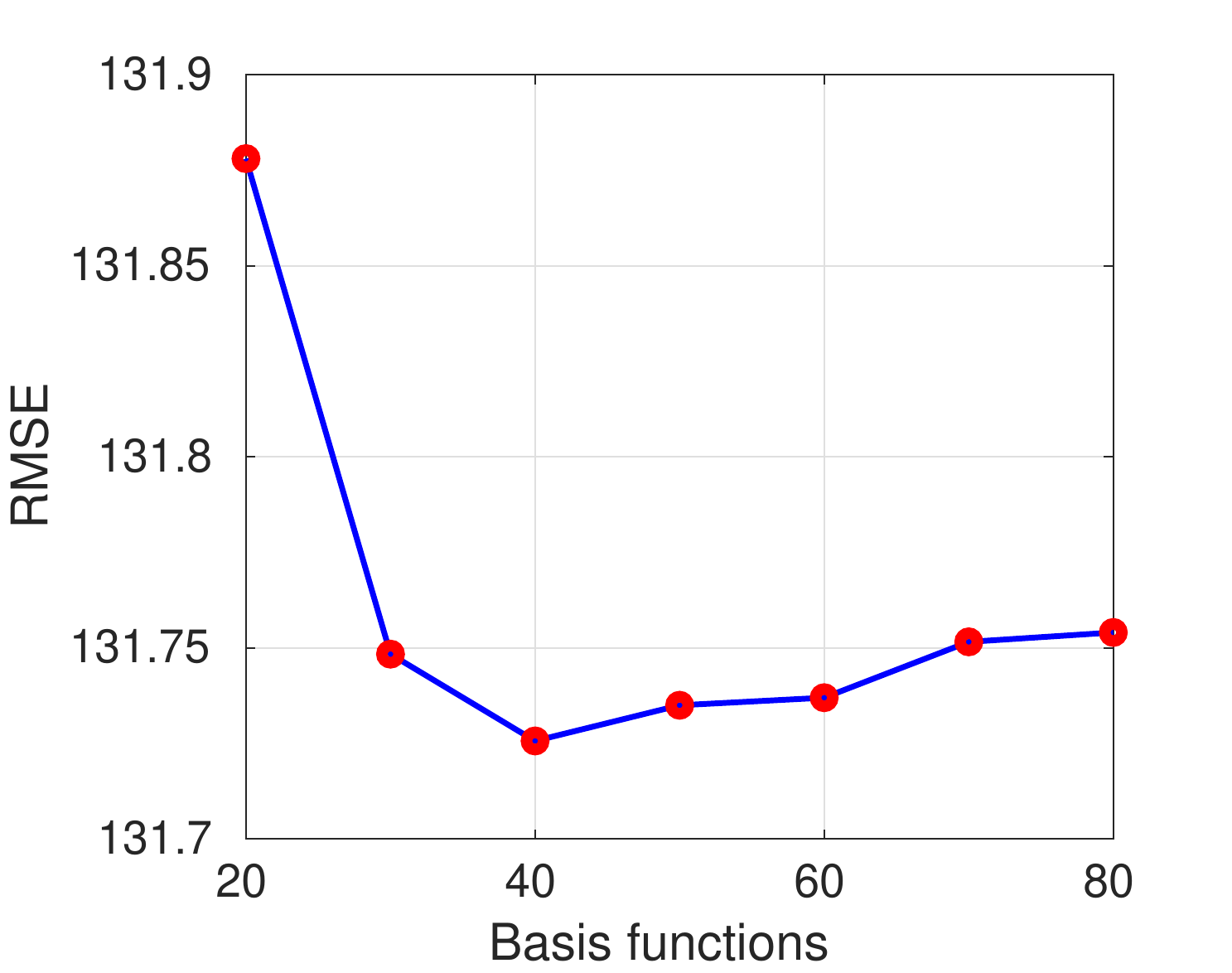}
    \label{Correlation21}
  }
  \subfigure[] {
    \includegraphics[width=0.23\textheight]{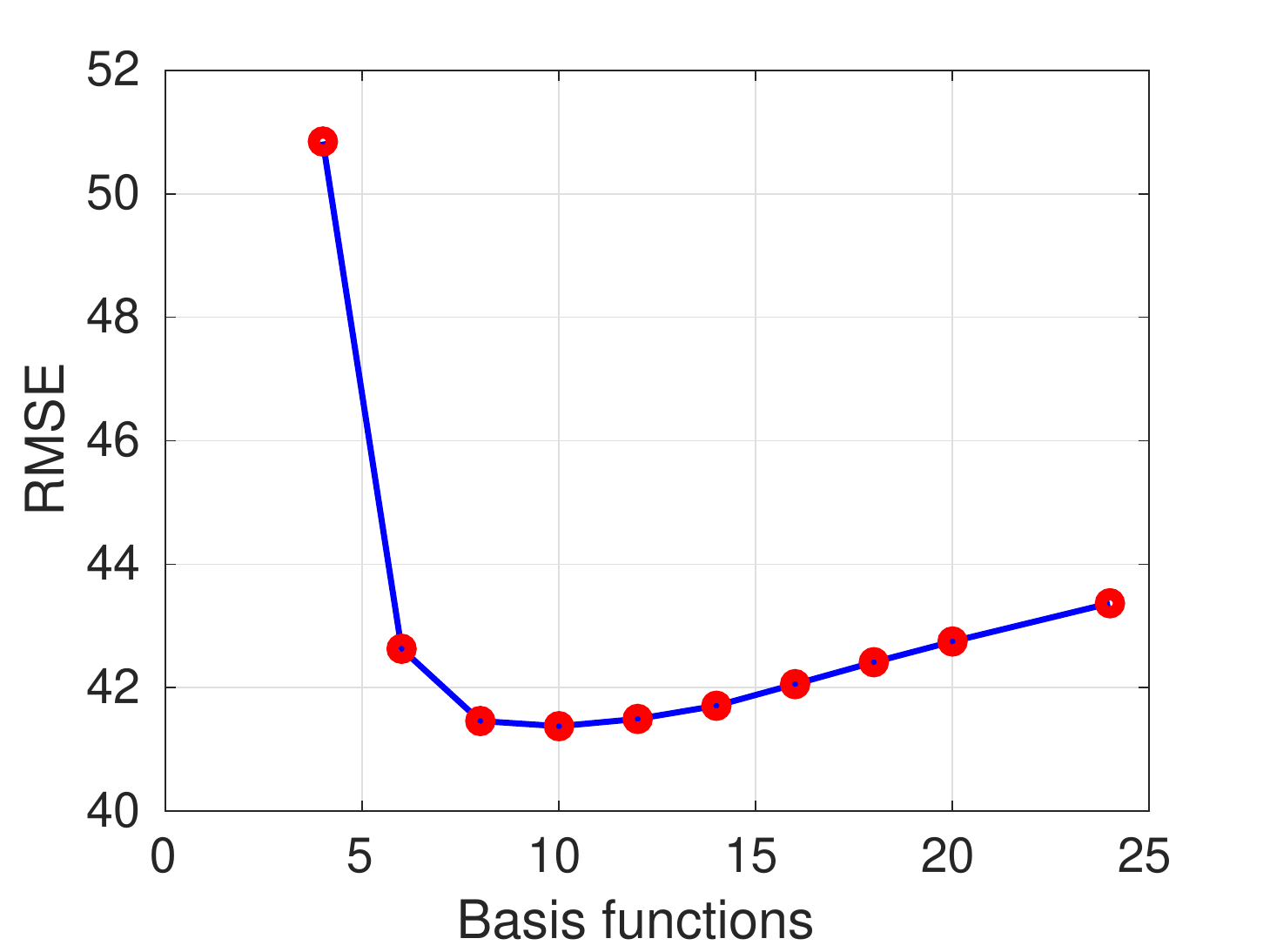}
    \label{Correlation33and43}
  }
  \subfigure[] {
    \includegraphics[width=0.23\textheight]{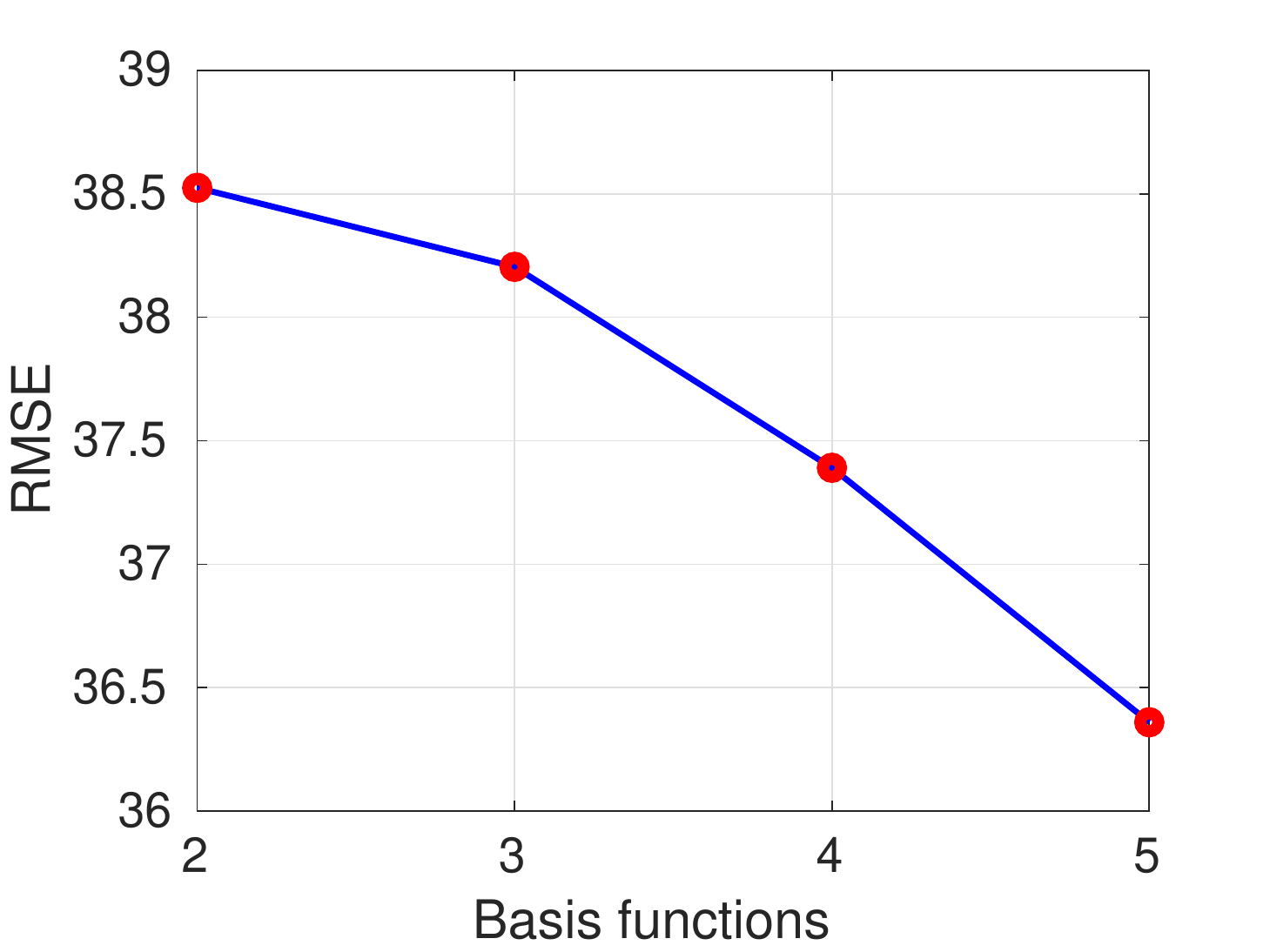}
    \label{Correlation42and51}
  }
  \caption{The optimum number of basis functions required to describe a
    correlation corresponds to the minimum value in the RMSE profile.
    Shown here are the variations in the RMSE scores with respect to the
    number of basis functions used to capture $\ref{Correlation21}$ two-body,
    $\ref{Correlation33and43}$ correlations with three bonds,
    and $\ref{Correlation42and51}$ correlations with four bonds.
    Note that the charge density values have not been scaled by the inverse of
    the supercell volume (i.e. by 23.5$^{3}\; \rm\AA^{3}$). Thus, the minima in
    these plots correspond to a RMSE of 0.01 (a), 3.19$\times
    10^{-3}$ (b) and 2.80$\times 10^{-3}$ (c) $e/\rm\AA^{3}$ in the predicted
    charge densities.}
  \label{BasisFunctionsNeeded}
\end{figure*}

Next, we calculate the number of basis functions needed to model the
two-body correlations. For this, we use 20-40 structures (equal number
of structures from each of the two sets in Section IIIA) and corresponding
ground state charge densities to train the model and another 20 structures
(10 each from the sets A and B in Section IIIA) for testing. The charge
density obtained from DFT calculations is represented using 
a 280$\times$280$\times$280 (supercell dimensions are 23.50$\times$23.50$
\times $23.50 $\rm\AA^{3}$) grid size. To reduce data redundancy, we use
charge density information from grid points separated by 10, 14, 20 or
28 grid points along each axes (i.e. along each dimension we use 280/10 =
28, 280/14 = 20, 280/20 = 14 and 280/28 = 10 grid points to
train our model, respectively).

For two-body correlations, Fig $\ref{Correlation21}$ shows the RMSE in
predicted values of the electronic density at 28$^{3} \times 20$
grid points (i.e. 28$^{3}$ grid points each from 20 charge density files)
from the test data set. To calculate the number of Chebyshev polynomials
needed to capture the two-body correlations, we truncate the model in
Eq. $\ref{ElectronicDensityLinearModel1}$ to the following
\begin{equation}
  \rho \left( {\bf r}_{ijk} \right) = \sum_{n = 0}^{N_{21}}
  c^{ \left( 21\right) }_{n} \left[ \sum_{p = 1}^{M_{ijk}} \psi_{n} \left(
    \left|{\bf r}_{ijk} - {\bf r}_{p} \right| \right) \right] = \sum_{n =
    0}^{N_{21}} c^{ \left( 21\right) }_{n} \phi^{ \left( 21\right) }_{n} 
\end{equation}
where, $M_{ijk}$ is the number of neighbors of grid point $\left( i, j, k
\right)$ and $N_{21}$ is the number of Chebyshev polynomials used to represent
the two-body correlations. The RMSE error in Fig $\ref{Correlation21}$
shows a marginal decrease as the number of basis functions increases and reaches
a minimum at 40 Chebyshev polynomials. Thus, we use 40 basis functions to
capture two-body correlations in the subsequent analysis. Using a similar
procedure, we see that the RMSE reaches a minimum when
Chebyshev polynomials of order 20 per bond are used to represent three-body
correlations with two bonds (see Fig. $\ref{Correlations3body}$, left). This
means that there are 21$\times \left(21+1\right)/2 =$ 231 correlation
coefficients for this type of three-body correlation.

\begin{figure}[htp]
  \centering
  \subfigure[] {
    \includegraphics[width=0.30\textheight]{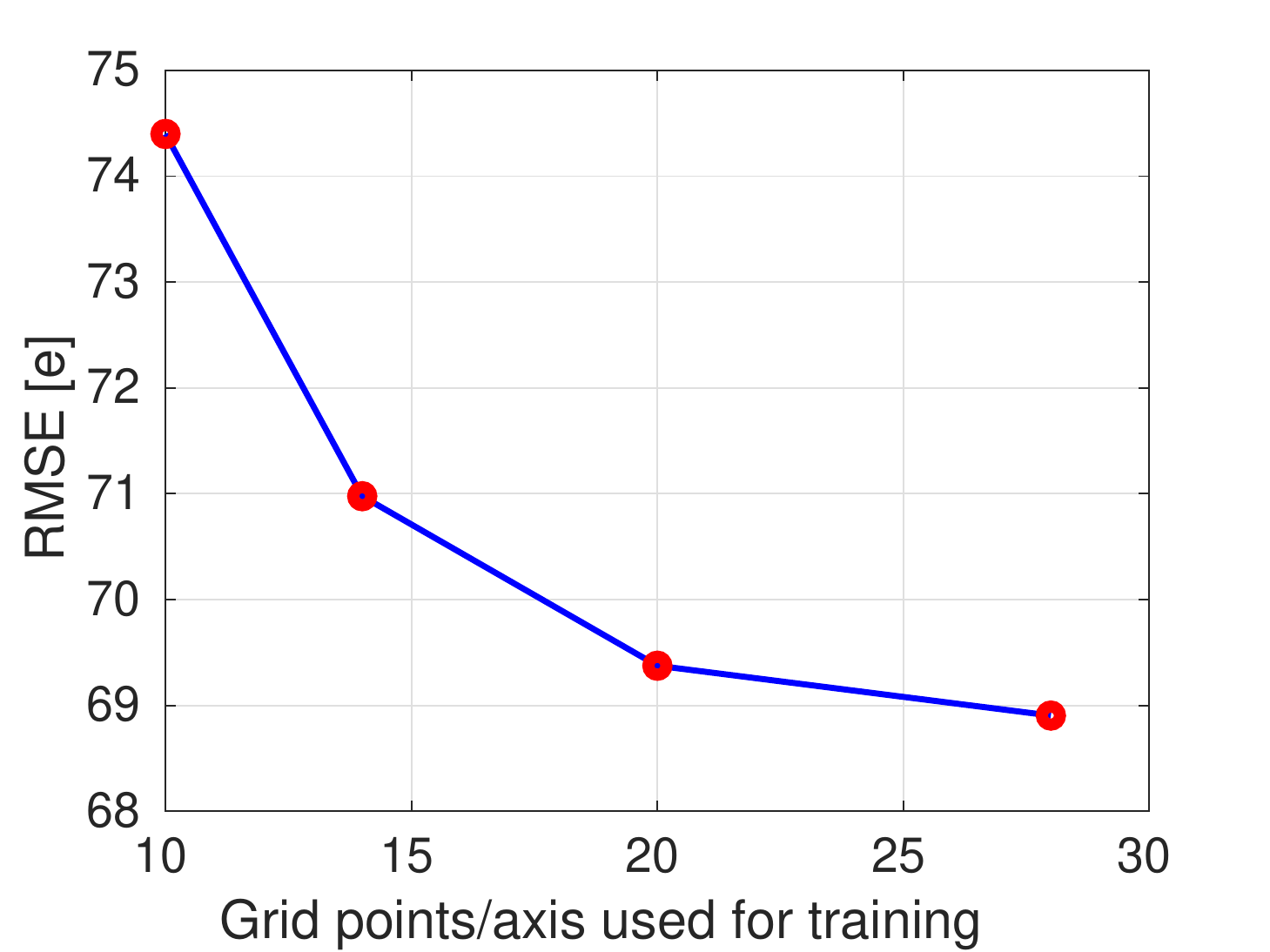}
    \label{number_of_grid_pts_convergence}
  }
  \subfigure[] {
    \includegraphics[width=0.30\textheight]{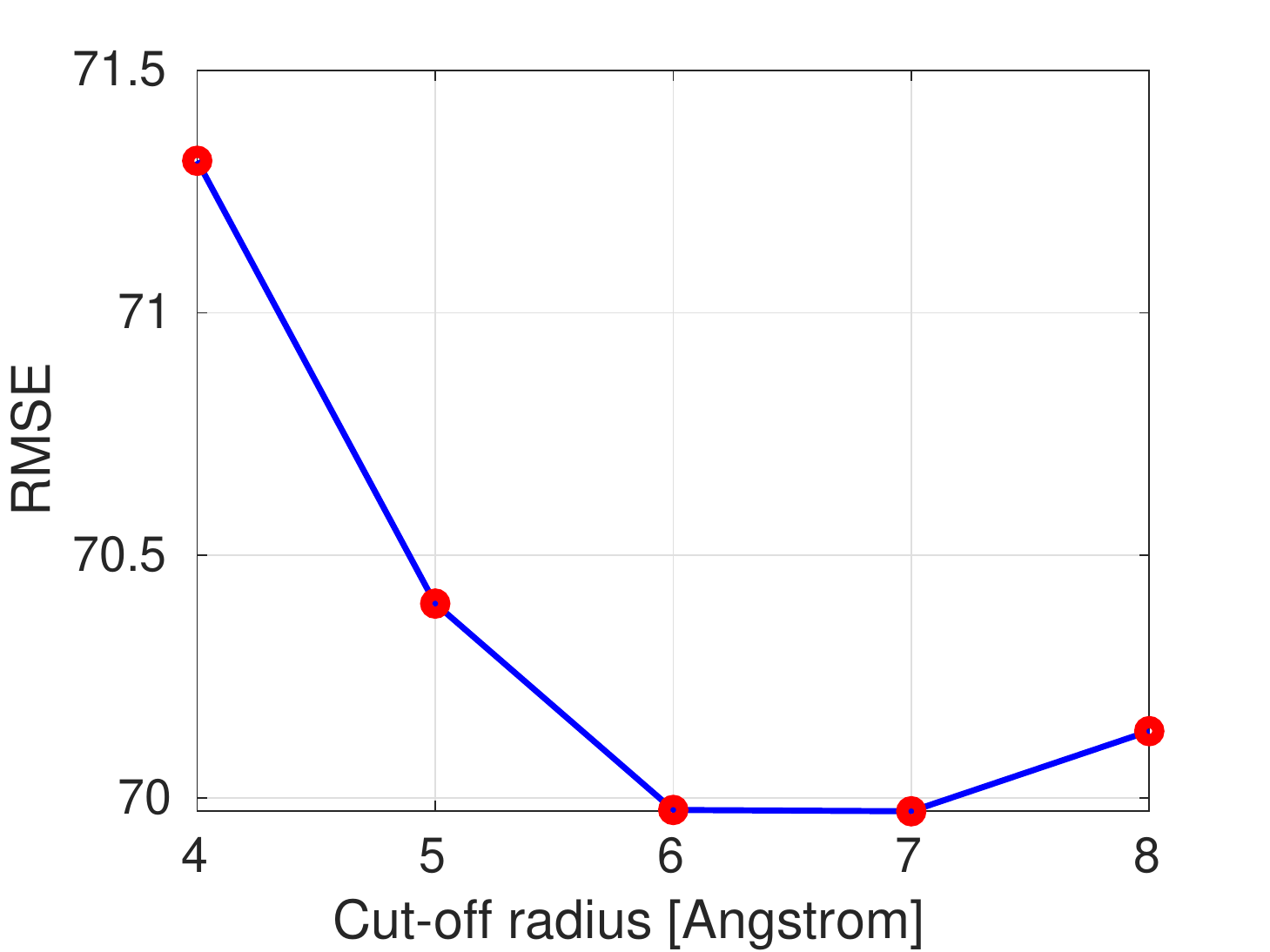}
    \label{cutoff_radius_convergence}
  }
  \caption{Shown here is the convergence of the RMSE with respect to
    the number of grid points along each edge vector and cut-off radius
    used to determine the number of neighbors surrounding an atom. Note
    that the charge density values are scaled by the inverse
    of the supercell volume (i.e. by 23.5$^{3}\; \rm\AA^{3}$). Thus, the
    minima in (a) and (b) correspond to RMSE of $\sim$5.4$\times
    10^{-3}$ $e/\rm\AA^{3}$}
  \label{Convergence_CutoffRadius_GridPoints}
\end{figure}

Figure $\ref{Convergence_CutoffRadius_GridPoints}$ shows the convergence
of the RMSE, for a model containing correlations with one- and two-bonds,
respect to a few parameters of the model. For example,
in Fig. $\ref{number_of_grid_pts_convergence}$ we see that the RMSE score
decreases with an increase in the number of grid points per charge density
file: The RMSE decreases from 74.39 (5.72$\times 10^{-3}$ e/$\rm\AA^{3}$)
to 68.90 (i.e. 5.30$\times 10^{-3}$ e/$\rm\AA^{3}$) when the number of grid
points increases from 10 grid points/axis (i.e. 1000 grid points per
charge density file) to 28 grid points/axis. Similarly,
Fig. $\ref{cutoff_radius_convergence}$ shows the convergence of the RMSE
with respect to the cut-off radius. % used to identify atoms that lie in the
%neighborhood of a grid point.
Thus, for the results shown in
Fig. $\ref{BasisFunctionsNeeded}$ we use 6.50 $\rm\AA$ as the cut-off
radius and this results in a RMSE of 5.4$\times 10^{-3}$ e/$\rm\AA^{3}$
(see Fig. $\ref{cutoff_radius_convergence}$).  
  
Figure $\ref{Correlation33and43}$ shows the number of basis functions per
bond needed to capture correlations with three-bonds, i.e. $c^{\left( 33
  \right)}$ and $c^{\left(43 \right)}$. Since both of these correlations
contain three-bonds, we assign the same number of Chebyshev polynomials
to both of them. Thus, for this analysis, the model in
Eq. $\ref{ElectronicDensityLinearModel1}$ is reduces to the following form
\begin{equation}
  \begin{split}
    &\rho \left( {\bf r}_{ijk} \right) = \sum_{n = 0}^{40} c^{
      \left( 21\right) }_{n} \phi^{ \left( 21\right) }_{n} +
    \sum_{n = 0}^{20} \sum_{m = n}^{20} c^{ \left( 32 \right) }_{mn}
    \phi^{ \left( 32\right) }_{mn} + \\
    &\sum_{n = 0}^{N_{33}} \sum_{m = n}^{N_{33}} \sum_{o = m}^{N_{33}}
    \left[ c_{mno}^{\left( 33 \right)} \phi_{mno}^{\left( 33\right)} + 
      c_{mno}^{\left( 43\right)} \phi_{mno}^{\left( 43\right)} \right]. 
  \end{split}
\end{equation}

From Fig. $\ref{Correlation33and43}$ we see that the RMSE score reaches
a minimum value when basis functions (per bond) of index 11 are used.
Thus, for all subsequent calculations, we use $2 \times \left(11
\times 12\times 13/6 \right) = 572$ correlation coefficients for both
$c_{mno}^{\left( 33 \right)}$ and $c_{mno}^{\left( 43\right)}$ correlations.
This means that a model that contains correlations with one, two, and
three bonds has in total $\left(40 + 231 + 572\right) = 843$ correlation
components and an equal number of correlation coefficients. Figure
$\ref{Correlation42and51}$ shows that the RMSE score decreases from a
minimum value of $3.19\times 10^{-3}$ e/$\rm\AA^{3}$ in
Fig. $\ref{Correlation33and43}$ to $2.92\times 10^{-3}$ e/$\rm\AA^{3}$
when correlations with four bonds are present in the model. For this
analysis, we use a model containing four-
and five-body correlations with four bonds:
\begin{equation}
  \begin{split}
    &\rho \left( {\bf r}_{ijk} \right) = \sum_{n = 0}^{40} c^{
      \left( 21\right) }_{n} \phi^{ \left( 21\right) }_{n} +
    \sum_{n = 0}^{20} \sum_{m = n}^{20} c^{ \left( 32 \right) }_{mn}
    \phi^{ \left( 32\right) }_{mn} + \\
    &\sum_{n = 0}^{10} \sum_{m = n}^{10} \sum_{o = m}^{10}
    \left[ c_{mno}^{\left( 33 \right)} \phi_{mno}^{\left( 33\right)} + 
      c_{mno}^{\left( 43\right)} \phi_{mno}^{\left( 43\right)} \right] + \\
    &\sum_{n = 0}^{N_{44}} \sum_{m = n}^{N_{44}} \sum_{o = m}^{N_{44}} \sum_{p = o}^{N_{44}}
    \left[ c_{mnop}^{\left( 44 \right)} \phi_{mnop}^{\left( 44\right)} + 
      c_{mnop}^{\left( 54\right)} \phi_{mnop}^{\left( 54\right)} \right].
  \end{split}
\end{equation}
The RMSE score decreases as more Chebyshev polynomials are included in the
model (see Fig. $\ref{Correlation42and51}$), but the computation cost for
evaluating these correlation components increases very quickly.
For example, when Chebyshev polynomials of index $n$ are included, the
number of correlation components is given by $\left(n + 1\right) \left(n
+ 2\right) \left(n + 2\right) \left(n + 3\right)/12$. Thus, we use
% have limited our analysis to
only a few basis functions to capture correlations with four bonds.

\begin{figure}[htp]
  \centering
  \subfigure[] {
    \includegraphics[width=0.30\textheight]{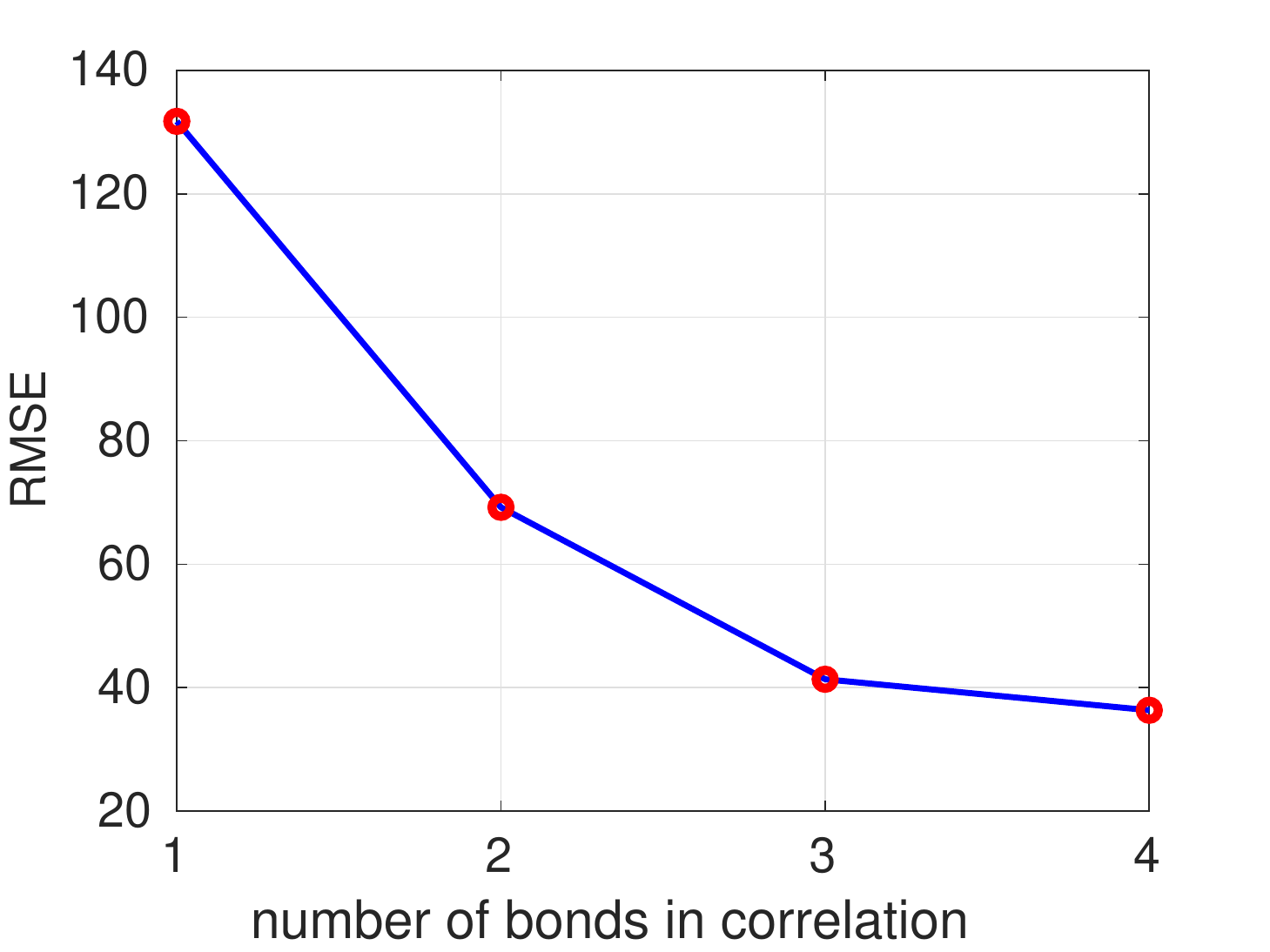}
    \label{error_correlation_type}
  }
  \subfigure[] {
    \includegraphics[width=0.30\textheight]{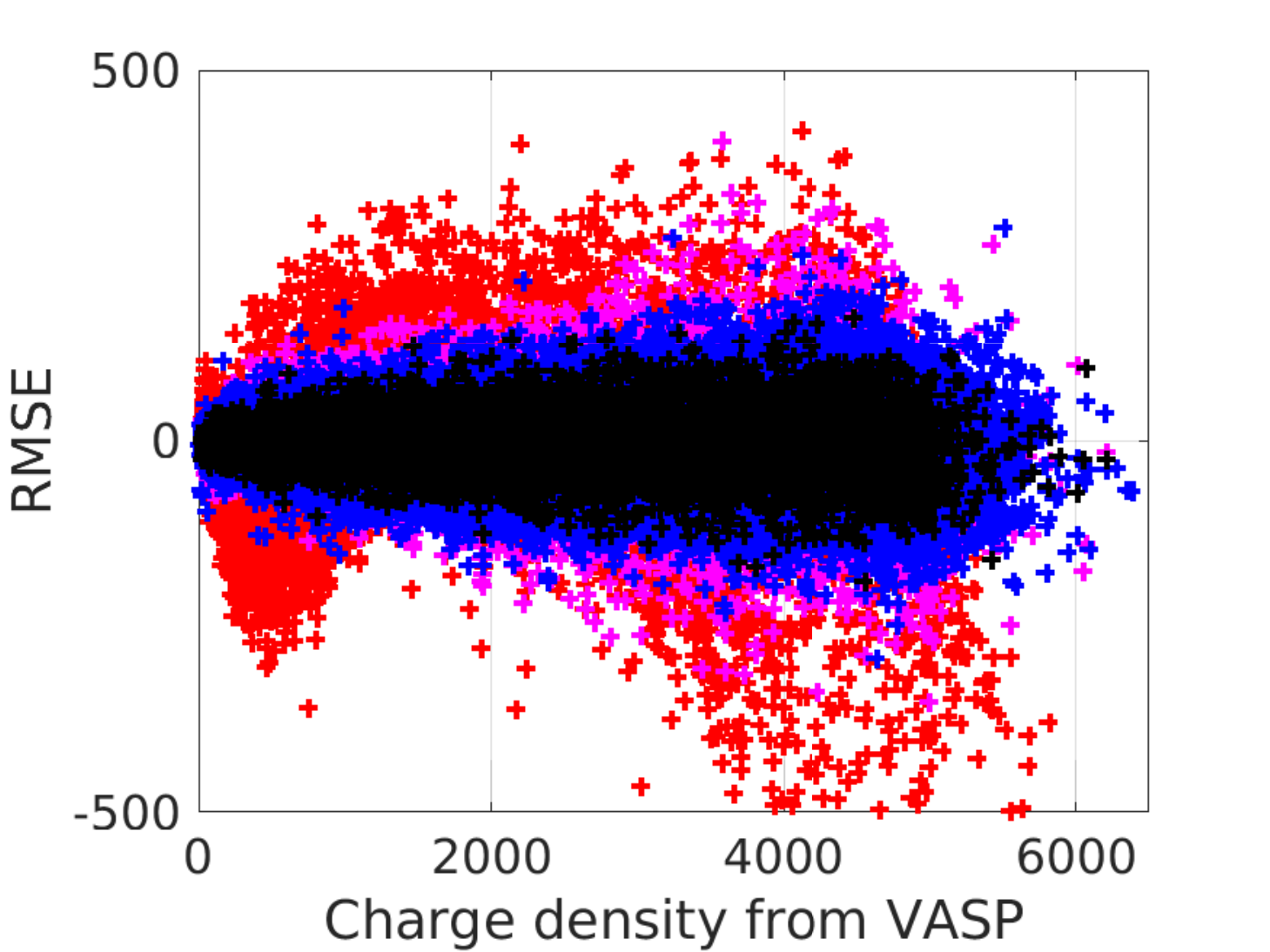}
    \label{model_error_comparison}
  }
  \caption{$\ref{error_correlation_type}$ shows the monotonous decrease
    in the RMSE with the increase in the order of correlations present in the
    model. $\ref{model_error_comparison}$ shows that predictions made
    using only two-body correlations (red markers) results in systematic
    errors in the predicted charge density. On the other hand, predictions
    made by using models containing correlations with two and three bonds
    (blue markers) do not exhibit systematic errors. Predictions made
    by using models containing correlations with four bonds significantly
    increases the accuracy. Note that the charge density values are not
    scaled by the inverse of the supercell volume (i.e. by 23.5$^{3}\;
    \rm\AA^{3}$). }
  \label{ErrorConvergence}
\end{figure}

Figure $\ref{error_correlation_type}$ shows the systematic increase in
the accuracy of predictions as the number of bonds in the correlation
increases. These results suggest that a model that includes correlations
with one, two and three bonds can very accurately capture the local
environment around a grid point and hence can to predict the ground
state electronic charge density. 

Figure $\ref{model_error_comparison}$ compares errors from models
containing different many-body terms. For example, predictions made,
at different grid points, by using only two-body correlations (red
markers) results in systematic errors at small as well as large values
of the charge density. On the other hand, predictions made (at different
grid points) by using models containing correlations with two and three
bonds (blue markers) do not exhibit systematic errors and predictions
made by using models containing correlations with four bonds (black
markers) have significantly higher accuracy.

\section{Results: Total Energy}
\label{results-energy-computation}

Next, to predict the total energy of an atomic configuration, we use
the coefficients of the many-body expansion (of the electronic density)
as descriptors, or features, to fit the total energy by using the parametric
and the non-parametric regression methods described in Section II. Thus,
Eq. $\ref{eq:linden1}$ is solved for each structure in the data set
and a coefficient vector (i.e. $\bar{\bf c}_{u}$) is obtained for
each structure (with index $s$). 

We conjecture that these coefficients are good descriptors since they
describe the amplitudes of each many-body contribution to the electronic
density. However, the number of coefficients also
depends on the number of Chebyshev functions necessary to capture
high-frequency density variations. We use 40 Chebyshev basis functions
to describe the two-body correlations, Chebyshev polynomials of index
20/bond to describe the correlations with two bonds and
Chebyshev polynomials of index 11/bond to describe the correlations with three-bonds.
This is based on the convergence analysis presented in Section IIIB. Thus,
the number of coefficients, or features, of the data set is 843. In the
subsequent sections, we present an analysis of the total energies
predicted using various linear and nonlinear regression techniques. 

To generate the model, we obtain training and validation sets by
selecting an equal number of structures from the two sets mentioned
in Section IIIA, i.e. %. Thus, for the subsequent analyses of different models,
we use 200 samples for training, another 200 samples for validation
and the remaining unseen samples are used to quantify the predictive
capabilities of the models. 
%to test the predictive
%capabilities of the models.
Validation is done using the three-way
hold-out method and the regularization coefficient is obtained by
minimizing the error using the Nelder-Mead algorithm.

\subsection{Linear Ridge Regression Model}
\label{linear-ridge-model}

\begin{figure}
\centering
\includegraphics[width=.80\linewidth]{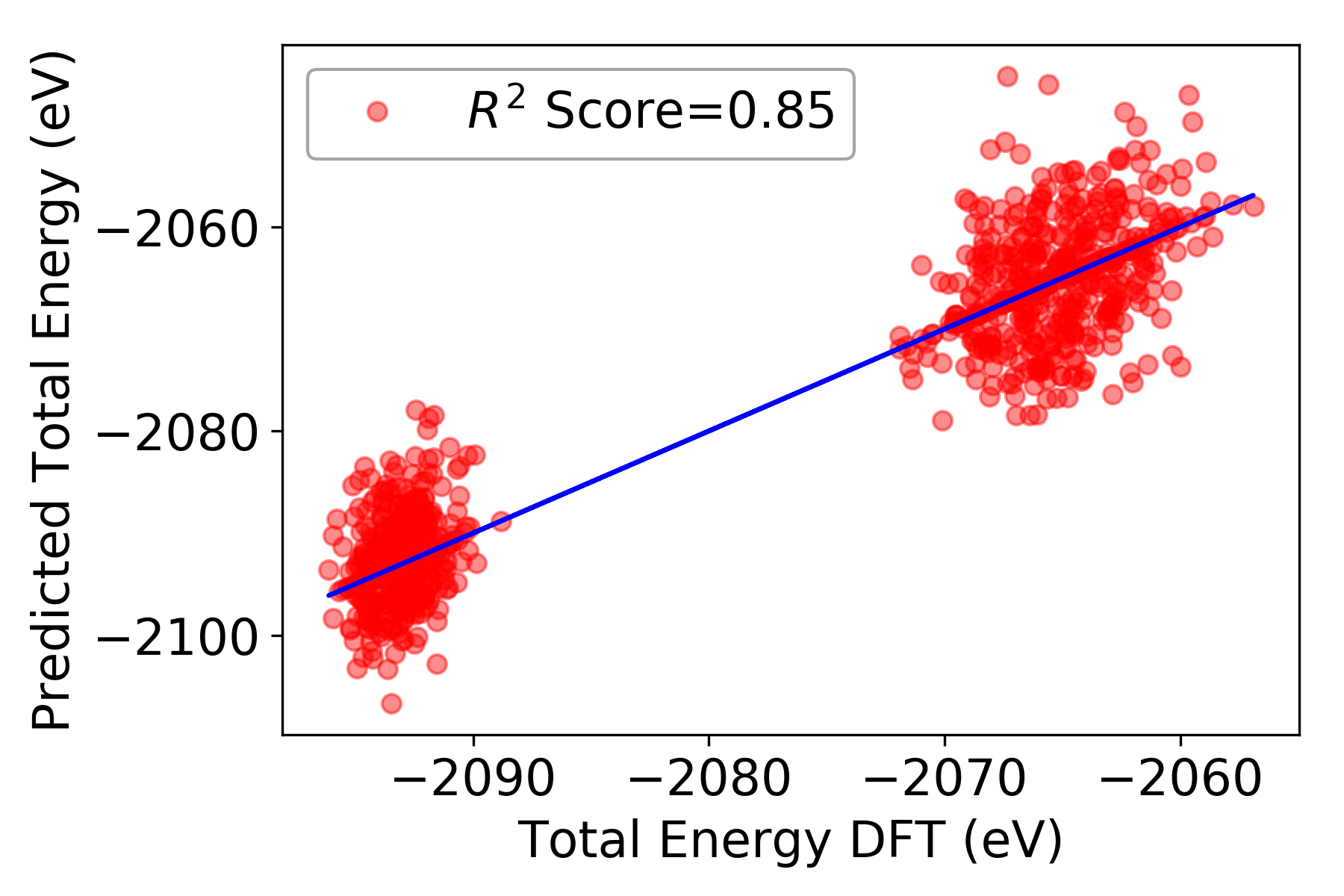}
\caption{Predicted total energies of structures in the validation set
  using features in the whole feature space. The two clusters correspond
  to structures in sets A and B (see Section $\ref{data_set_description}$).}
\label{fig:linearce}
\end{figure}

To model the total energy of a configuration using Eq. $\ref{eq:linener}$,
we use descriptors (i.e. coefficient vectors $\bar{\bf c}^{\left( s\right)}$)
in the full feature space and optimize the $R^{2}$ score to obtain
the optimum value of the regularization coefficient $\lambda$ in Eq. $\ref{eq:linener}$.
%The optimum value of the regularization constant was {\color{blue}found to be $\lambda^{\ast}
%= XXXX$. Using this $\lambda^{\ast}$, Fig. XX shows}
%the distribution of coefficients, i.e. components of the $\beta$ vector.
The total energies of structures in the validation data
set obtained from this $\beta$ is shown in $\ref{fig:linearce}$. The solid
line in Fig. $\ref{fig:linearce}$ corresponds to a scenario where
the predictions made from a model agrees perfectly with the DFT-calculated
total energies. As explained in Section IIIA, there are two
clusters in our data set and they correspond to different magnitudes
of perturbation of atomic positions. In Fig. $\ref{fig:linearce}$,
total energies predicted for structures in data set A (i.e. the lower-left
cluster) show a smaller variance compared to the predictions made
for the structures in data set B (i.e. the upper-right cluster). %The inset
%in Fig. $\ref{fig:linearce}$ shows that the model errors are not centered
%around zero. The angle that this distribution makes with the 
%horizontal axis is probably an indication of the systematic bias in
%the predictions. 

It is clear that these results correspond to an over-fitted regime: The
variance in the predicted energies for the input data set is smaller than
the variance in the predicted energies for the unseen data. 
%{\color{blue}Also, it is noticeable over-fitting of the training set and large deviations of the
%  predicted data.}
In principal, to reach a better bias-variance trade off, the model
needs more data to learn the features properly. This issue can be
tackled by simply increasing the amount of training samples. However, our
goal here is to minimize the overall variance as well as the number of
training samples. Thus, an alternative approach to achieve this goal is
to reduce the model complexity by feature selection or dimensionality
reduction. For this purpose, we use the principal component analysis
(PCA) method to reduce the dimension of the features space and reach a
better model performance in which 200 samples are used for training,
another 200 samples are used for validation, and the remaining unseen
data are used for testing.

\subsection{Dimensionality reduction using principal component analysis}
\label{pca-dimenstionality-reduction}

\begin{figure}
\centering
\includegraphics[width=0.80\linewidth]{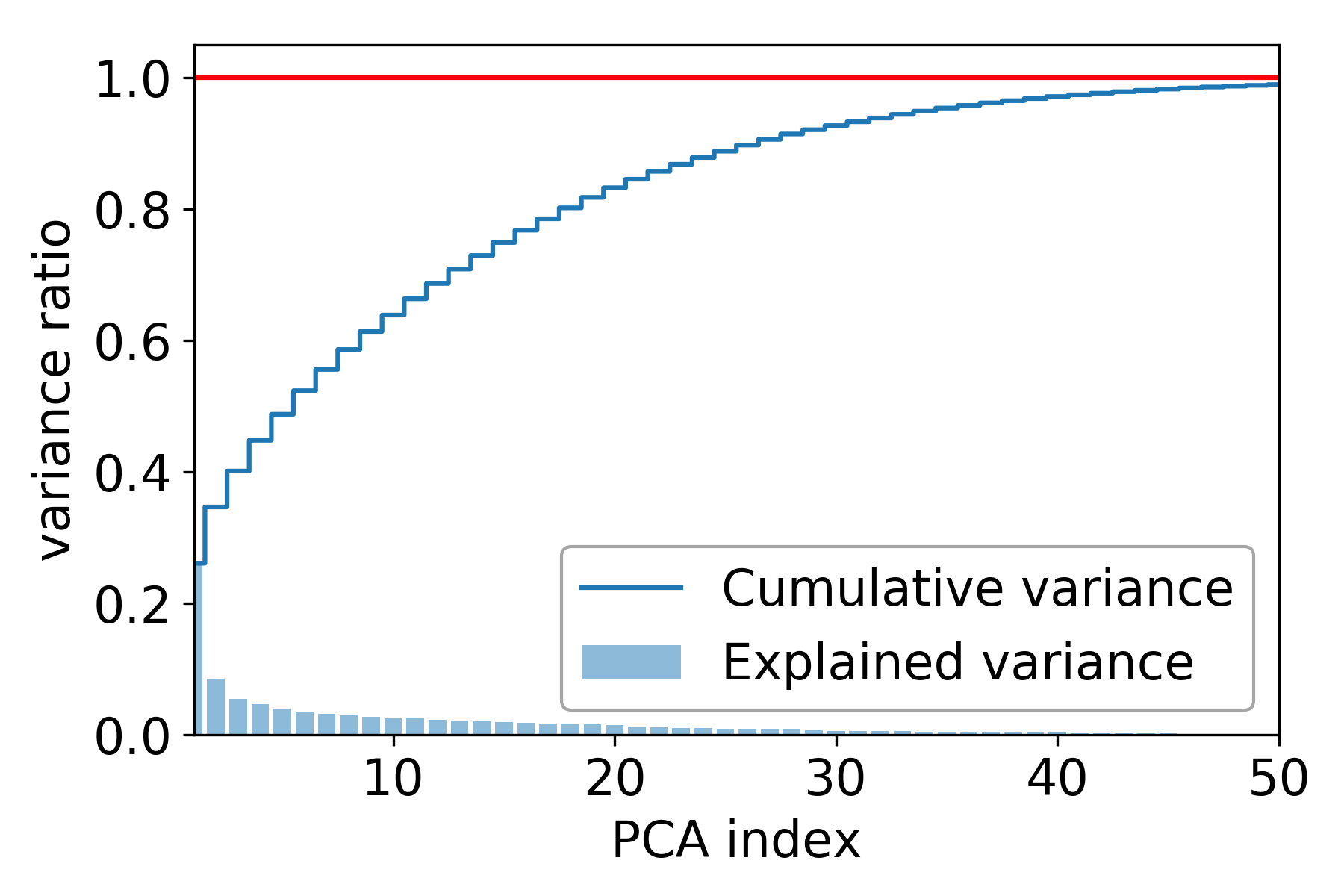}
\caption{The contributions of the different principal components to the
  total variance is shown using the solid line. The solid bars denote
  the magnitude of the normalized eigenvalue of along each principal
  components.}
\label{fig:varexp}
\end{figure}

To reduce the dimensionality of the feature space, we use PCA on the set
of coefficient vectors $\{\bar{\bf c}^{\left(1\right)}, \bar{\bf c}^{ \left(
  2\right)}, \cdots  \}$ (solutions of the linear system in Eq.
$\ref{eq:linden1}$ for different structures and the corresponding
charge densities in the training data set). Figure \ref{fig:varexp}
shows the magnitude of the normalized eigenvalues arranged in descending
order. These normalized eigenvalues can be interpreted as the magnitude of
the variance of the
corresponding eigenvector, i.e. the largest eigenvalue encodes more variance
as explained in the methods section.

Figure \ref{fig:varexp} also displays the cumulative variance - interestingly,
using only 20 principal components our model can capture 80\% of the total
variance in the data.
%80\% of the total variance of the data was achieved using only 20 principal
%components.
In addition, as we systematically increase the number of principal
components, the cumulative sum quickly approaches 1 with only 50
normalized eigenvalues which is just 6\% of the size of the feature
space. Thus, PCA helps us to select only a few relevant features, thereby
reducing model complexity (the number of principal components
$\ll \rm{dim}\left( \bar{\rm c} \right)$, i.e. the size of the feature
space). In addition, PCA avoids collinearity among the features by using
principal components that are, by design, orthogonal to each other.

After projecting the coefficient vectors, i.e. the descriptors, into the
important principal components, we use linear and nonlinear regression
techniques to train our model. This decrease in dimension of the feature
space (due to PCA) allows us to use a smaller data set for training these
models resulting in a better bias-variance trade off. 

\subsection{Principal components regression}
\label{principal-components-regression}

\begin{figure}
\centering
\includegraphics[width=0.80\linewidth]{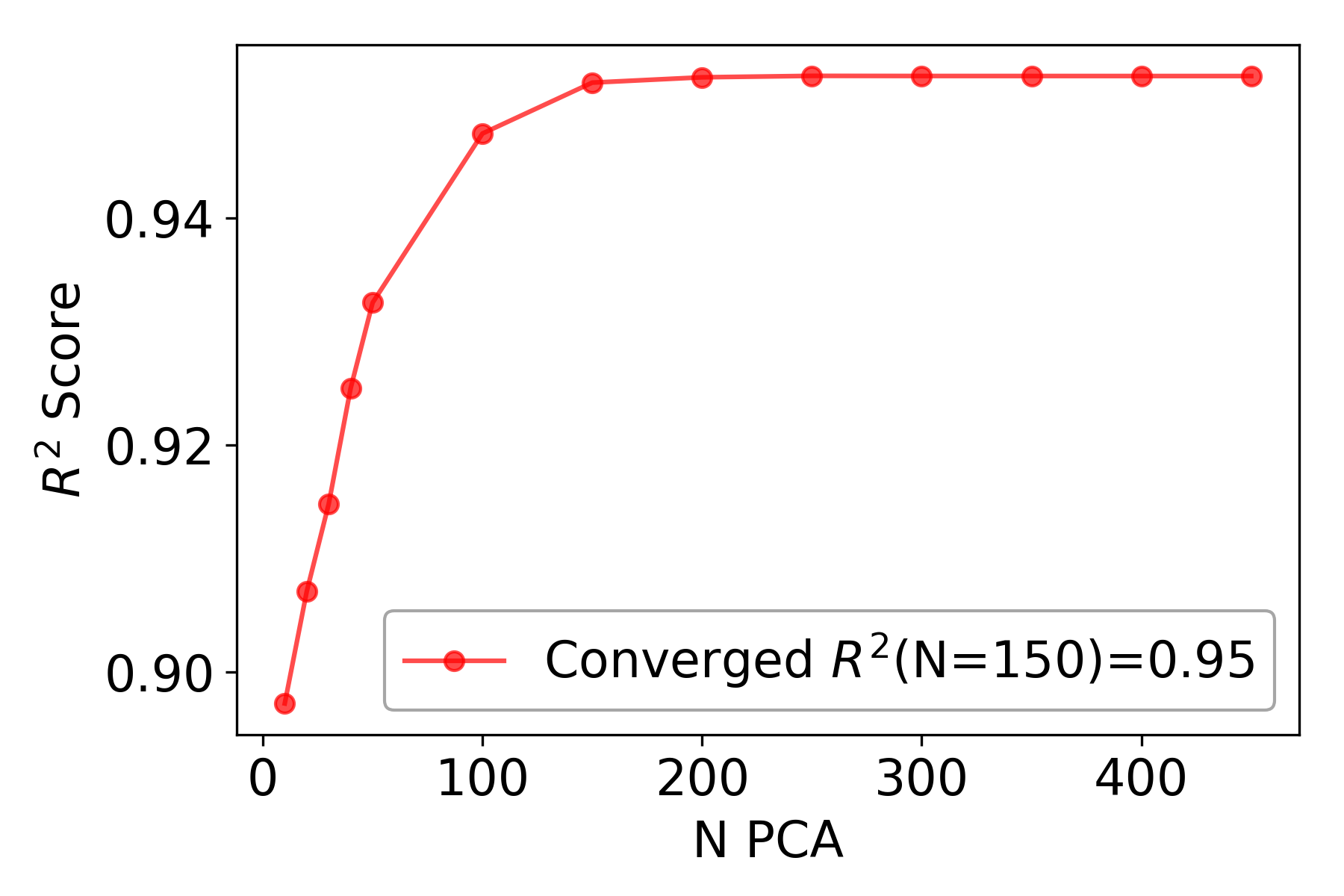}
\caption{Shown here is the convergence of the $R^{2}$ scores of the predicted
  total energies using a ridge regression model as a function of the number
  of principal components.}
\label{fig:pcrconv}
\end{figure}

Next, we use linear ridge regression on the PCA-projected
data. For this, we need to identify the minimum number of principal
components required to converge the $R^{2}$ score of the model. Figure \ref{fig:pcrconv}
illustrates the convergence of the \(R^2\) score with respect to the number
of principal components. It is clear that by using only ten principal
components we can capture 90 \% of the variation. Nonetheless, by
increasing the number of principal components to 150, we obtain
a model that covers 96\% of the variation.

\begin{figure}
\centering
\includegraphics[width=0.80\linewidth]{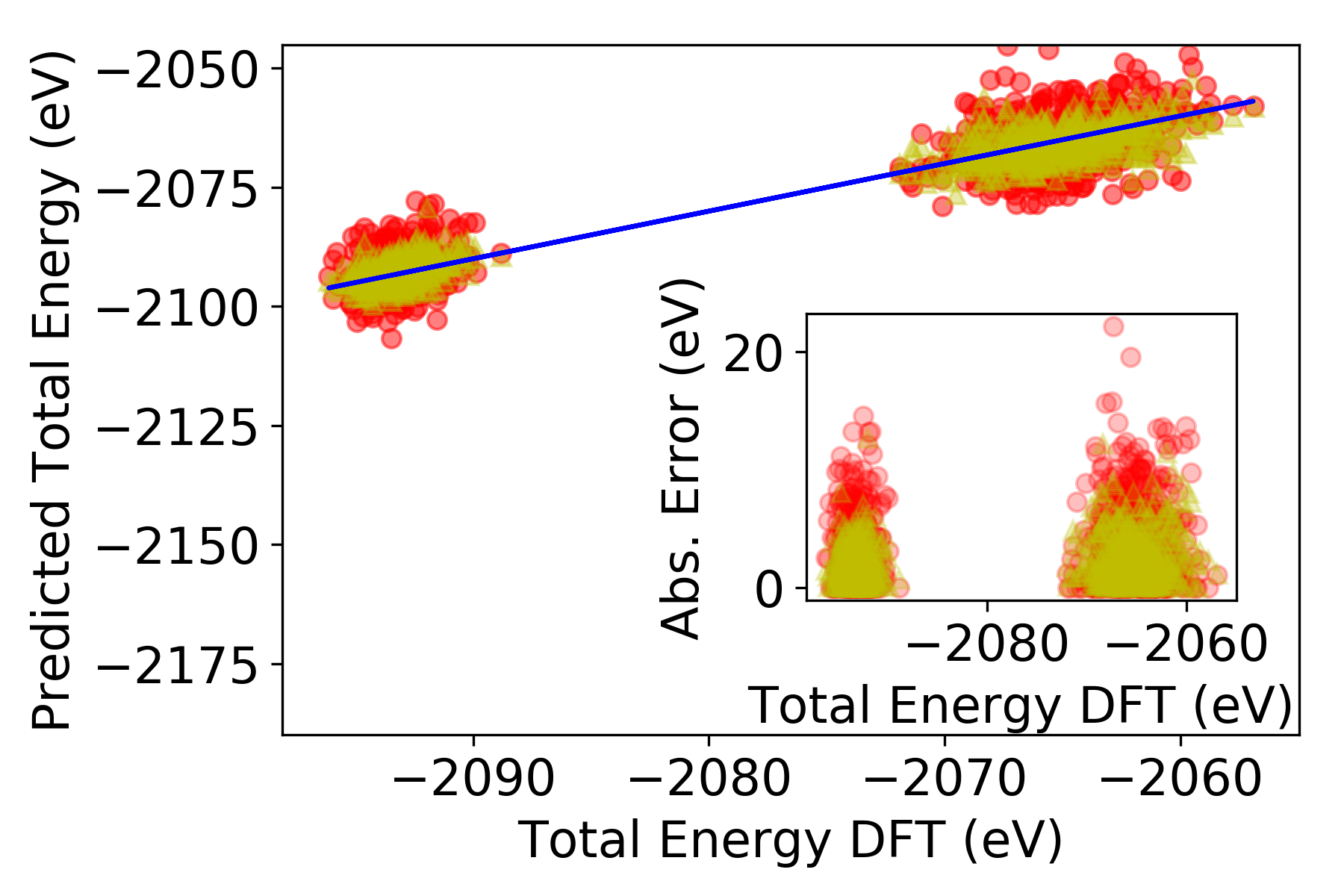}
\caption{Shown here is a comparison between the DFT calculated total
  energies and the total energies predicted using two ridge regression
  models: one that uses descriptors in the whole feature space (red
  circles), and another that uses descriptors projected on 150 principal
  components (triangles). The solid line (blue) corresponds
    to perfect agreement with the DFT data.}
\label{fig:pcrce}
\end{figure}

Figure \ref{fig:pcrce} compares the predictions made by using two different linear
regression models. The first model uses descriptor vectors %(i.e. coefficient
% vectors obtained by fitting each charge density separately)
in entire feature
space while the second model uses descriptor vectors projected into a
space of 150 principal components. The advantage of using principal
component regression is evident from the fact that by using only 150 principal
components and total energies of 200 training structures we are able to
significantly improve the performance of the model. In addition, the fact that
by using only 150 principal components our model achieves a $R^2$ score of 96\%
illustrates that deviations in the predicted values have diminished significantly.
Thus, principal component regression is able to capture important
features of the model. %To further improve the model predictions, we used
%kernel-based methods and compared their behavior with linear ridge regression. 

\subsection{Nonlinear models}
\label{kernel-methods-models-and-overall-comparison}

Next, we explore the nonlinear characteristics of the descriptors using
kernel-based methods like kernel ridge regression and Gaussian process
regression. To this end, we use a Gaussian kernel and descriptors projected
into the space of a few selected principal components. 

\begin{figure}
  \centering
  \includegraphics[width=.80\linewidth]{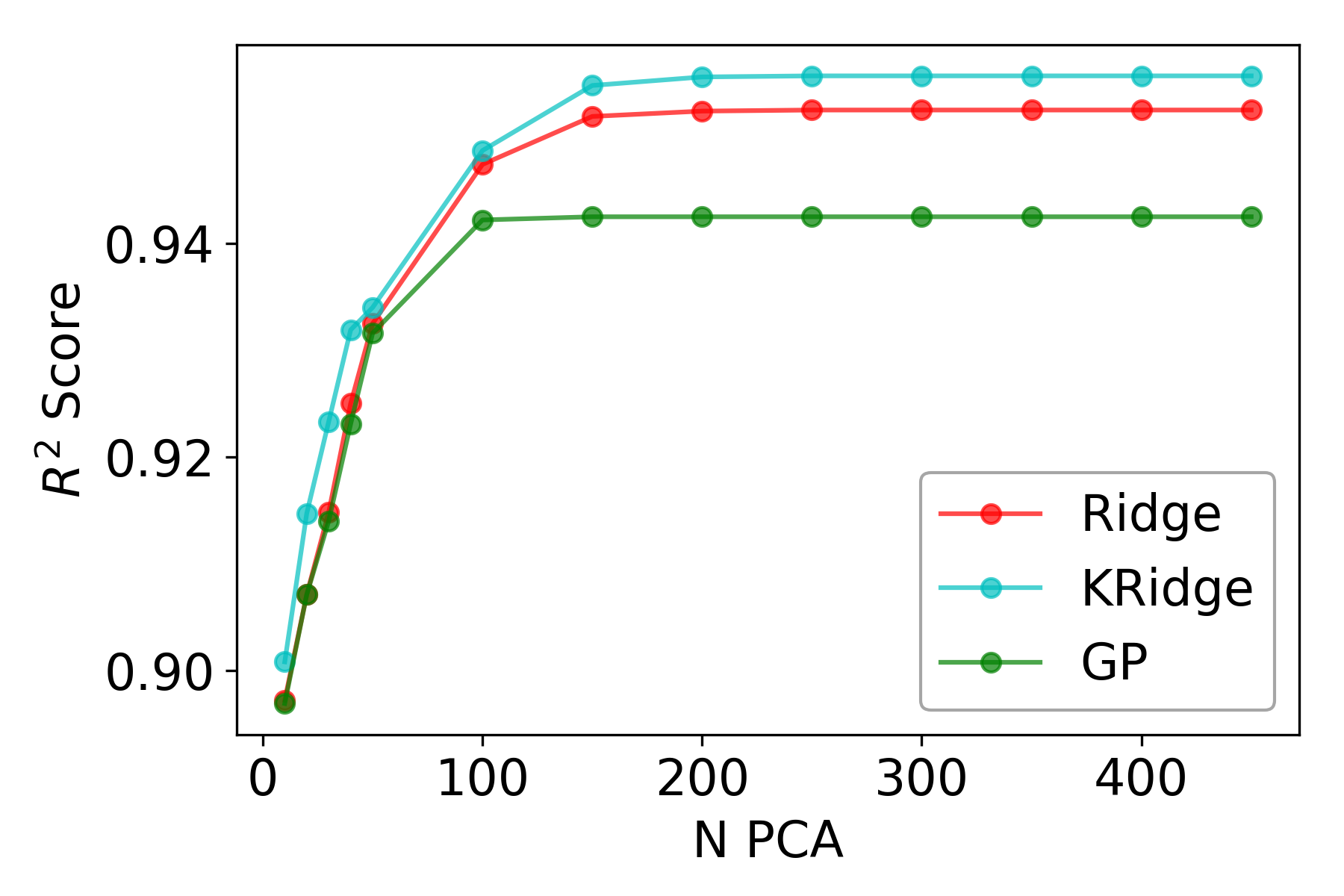}
  \caption{Shown here is the convergence of the $R^{2}$ scores (for the
    three different nonlinear regression techniques) with respect to the
    number of principal components used.}
  \label{fig:allconv}
\end{figure}

Figure \ref{fig:allconv} displays the convergence of the \(R^2\) score of
these kernel methods and compares them to predictions made using the principal
component regression method presented in the previous section. It is easy
to see that the linear and kernel ridge regression models exhibit similar
convergence behavior and their respective $R^2$ scores converges when
$\sim$150 principal components are present in the models. In addition, the
difference in their converged $R^2$ scores (in Fig. $\ref{fig:allconv}$) is less than
1\%, meaning that these two models capture similar information from the
training data. From Fig. $\ref{fig:allconv}$, we see that even though the
$R^2$ score of the Gaussian process regression model is similar to the
kernel and linear ridge regression models, it converges to a value that is
about 3\% less that these two models.

\begin{figure}
  \centering
  \includegraphics[width=.80\linewidth]{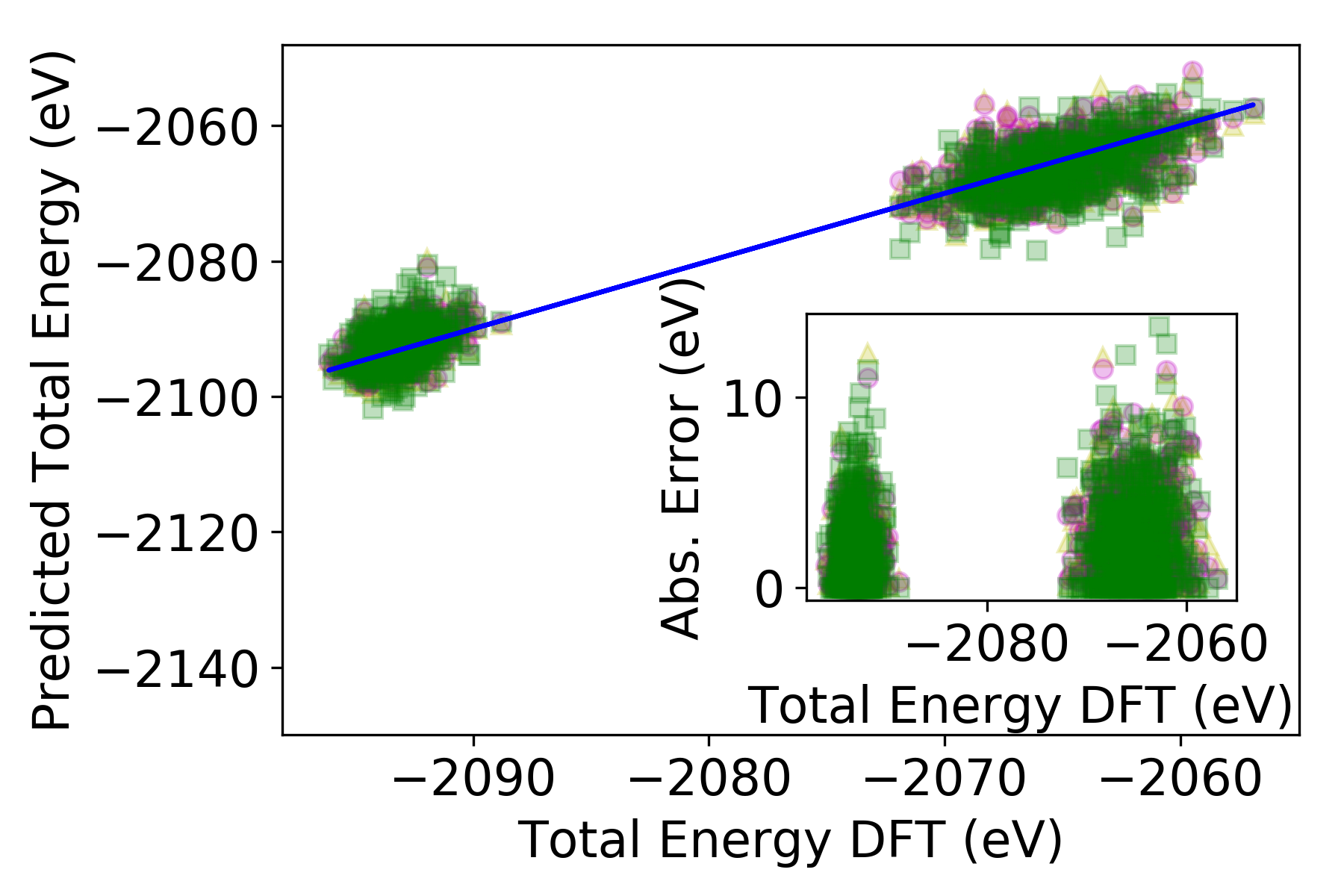}
  \caption{A comparison of the DFT calculated and predicted total energies
    for models obtained by using linear (triangles), Gaussian process
    (squares), and kernel ridge (circles) regression techniques. The solid
    line (blue) corresponds to perfect agreement with the DFT data.}
  \label{fig:allce}
\end{figure}

Figure \ref{fig:allce} compares the total energies predicted using
both linear and nonlinear regression models. Finally, we analyze the
performance of our models using several statistical measures, i.e.
the $R^{2}$ score, the mean squared error (MSE) and the mean absolute
error (MAE) for all the structures in the two data sets described in
Section IIIA. These values are listed in Table \ref{fig:tableae}.

\begin{table}
  \begin{tabular}{c|c|c|c}
    \hline
    \hline
    & Ridge & KRidge & GP \\
    \hline
    $R^{2}$ & 0.9784 & 0.9784 & 0.9753\\
    RMS (eV) & 2.909 & 2.7876 & 3.1084 \\
    MAE(eV) & 2.1531 & 2.0176 & 2.1648 \\
    \hline
    \hline
  \end{tabular}
  \caption{Performance of the linear and nonlinear regression models
    for all structures in the date set.}
  \label{fig:tableae}
\end{table}

\begin{figure}
  \centering
  \includegraphics[width=0.90\linewidth]{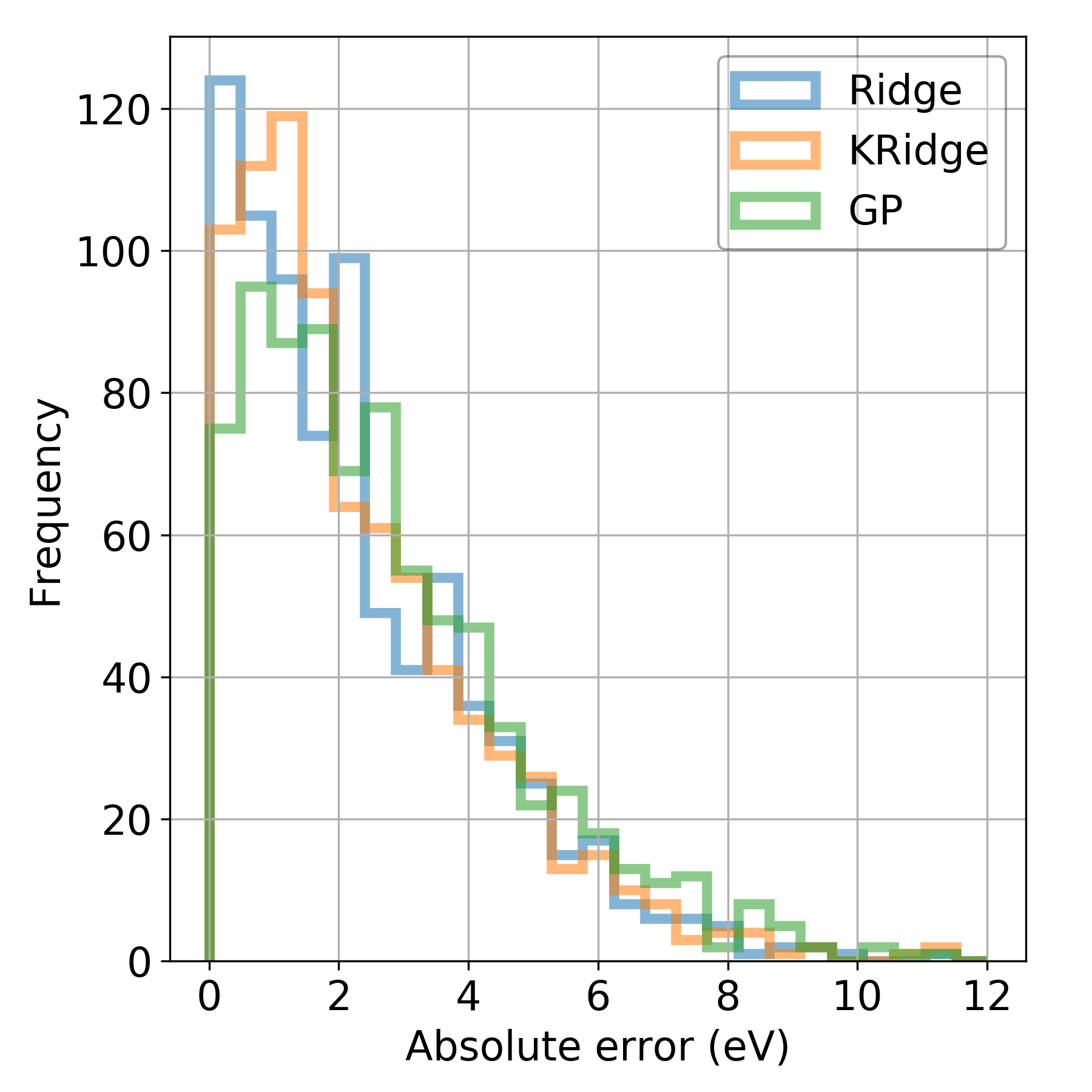}
  \caption{Shown here is the distribution of the absolute errors in the
    predicted total energies for all structures in the data set using
    ridge, kernel and GP regression techniques.}
  \label{fig:aec}
\end{figure}

Figure \ref{fig:aec} shows that the histograms of the absolute errors
in the predicted total energies for each configuration (in the whole
data set) using three different models exhibit similar trends. Thus,
from a statistical point of view, these %results suggest that the
three models offer equivalent performance.
Table \ref{fig:tableaepa} also compares the performance of the
models using different metrics: mean error, maximum positive
and negative errors per supercell and per atom. Thus, the maximum absolute
errors are 12.31, 11.50 and 13.71 eV/supercell for ridge, kernel ridge
and Gaussian process regressions, respectively meaning that the maximum
absolute error per atom is 0.027 eV which is less than the chemical
accuracy (1 kcal/mol = 0.043 eV/atom). 

\begin{table}
  \begin{tabular}{c|c|c|c}
    \hline
    \hline
    Measure/Method & Ridge & KRR & GPR \\
    \hline
    maximum negative & -8.09 & -9.26 & -10.24 \\
    error/supercell  &  &  &  \\
    \hline
    maximum positive & 9.82 & 9.20 & 12.31 \\
    error/supercell  &  &  &  \\
    \hline
    mean error/atom & 0.03 & 0.02 & 0.03 \\
    \hline
    maximum negative & -0.017 & -0.018 & -0.020 \\
    error/atom &   &  &  \\
    \hline
    maximum positive & 0.019 & 0.018 & 0.024 \\
    error/atom &  & &  \\
    \hline
    \hline
  \end{tabular}
  \caption{A comparison of the performance of models obtained by
    using three nonlinear regression methods using different metrics.}
  \label{fig:tableaepa}
\end{table}

\section{Summary and discussions}
\label{section_discussion}

Our results show that descriptors based on many-body correlations can
effectively capture local neighborhood information. For our analysis,
these many-body descriptors are calculated using a set of Chebyshev
polynomials (and their products), but any other set of orthogonal
polynomials can also be used. Using these descriptors we are able to reproduce
ground state electronic charge densities of amorphous structures to
a high level of accuracy. It is possible to systematically improve the
accuracy by incorporating higher order many-body correlations. In addition,
we are able to generate reliable models using only a few tens of structures
and their ground state charge densities. This is much smaller
than the hundreds of structures required to train some of the other
models presented in the literature.

Many other descriptors have been proposed in the literature: for example,
Brockherde et al. used cosine functions to represent the ground state
charge density, while Grisafi et al. used atom-centered symmetry-adapted
basis function based on spherical harmonics.\cite{GrisafiACS2019,
  GrisafiWilkinsCsyaniCeritti} However, from the cross-validation based
testing approach presented in Section IIIB, it
seems that the accuracy of a model containing a particular type of correlation
cannot be systematically improved by simply increasing the number of
orthogonal basis functions. This is because, beyond a certain value of
the RMSE score, increasing the number of basis functions can lead to
over-fitting. On the other hand, by incorporating two-, three- and four-body
correlations, we are able to systematically improve the predictive capability
of our models.

The role played by different many-body correlations may not be easily evident
in some of the popularly used descriptors in the literature. For example, the
bi-spectrum basis functions used in Ref. [\onlinecite{GrisafiACS2019}] can
capture up to four body correlations (with three bonds), but these correlations
correspond to $c^{\left(32\right)}$ (shown in Fig. 1(left)) and $c^{\left(43
  \right)}$ (shown in Fig. 2 (first column, left)) and do not include correlations
with closed loops (such as $c^{\left( 33 \right)}$. We see that correlations
with closed loops, like $c^{\left(33\right)}$, $c^{\left(44\right)}$, ...,
$c^{\left(46\right)}$, are more effective than correlations without closed
loops in enhancing the model performance. 
Here, we note that the importance of systematically incorporating many-body
correlations can
also be understood from the fact that simple metals (like Cu, Ag, Al, Au)
have been successfully modeled using embedded-atom-method type of interatomic
potentials (containing only two-body interactions and their products), but
three-body interactions are needed to model elements like Si and Ge.

Amplitudes of these many-body correlations are used to model the total
energy from the ground state charge density within chemical accuracy. These
amplitudes are invariant to global rotation/translation and permutation
of charge density grid indices and hence are useful to uniquely capture 
the distribution of electronic density corresponding to a distribution
of atoms. In the future, we plan to further explore model reduction using
compressed sensing to decrease the number of training samples required
to optimize the linear system in Eq. $\ref{eq:linener}$.

Brockherde et al.\cite{Brockherde2017} were able to reproduce total energies
of small molecules, like H$_{2}$ and H$_{2}$O, with high accuracy (mean
average error $\sim 0.01$ kcal/mol) using only 10 and 20 structures, respectively.
%, to train the respective models (mean average error $\sim 0.01$ kcal/mol).
But, for systems, like benzene and ethane (that contain more atoms than
that present in H$_{2}$ and H$_{2}$O), the number of structures required
to achieve a mean average error of 0.37 kcal/mol increased by two orders
of magnitude. Perhaps this is an indication that the local electronic
density around an atom can be used accurately map the energy of the atom,
and the sum of these energies can be used to accurately calculate the total
energy of the system. But, descriptors based on the total electronic density
of the system (containing as much as 512 atoms in our case) can perhaps
capture many important global characteristic features of the system, but
fail to capture salient details of $\rho \left( {\bf r} \right)$ close to
bond centers. Thus, in future, we plan to explore a framework to 
model the total energy as sum of energies of individual atoms and these atomic
energies can be obtained from the charge density in the vicinity of the atom.

\section*{Acknowledgments}

The authors wish to thank John Klepeis, Lorin Benedict, Arthur Tamm,
ShinYoung Kang, Seung Ho Hahn and Chiraag Nataraj for numerous stimulating
discussions. A. S. wishes to thank Mark Tuckerman for valuable suggestions
and encouragement to work on this problem during his visit to LLNL.
This work was performed under the auspices of the U.S. Department of
Energy by Lawrence Livermore National Laboratory under Contract
DE-AC52-07NA27344. Computing support for this work came from the Lawrence
Livermore National Laboratory (LLNL) Institutional Computing Grand
Challenge program. 

\section*{Data availability}
The data that support the findings of this study are available from
the corresponding author upon reasonable request.

% \bibliography{MyBibliography.bib}
\end{document}